\documentclass[aps,epsf]{iopart}
\usepackage{graphics}
\usepackage{graphicx}
\usepackage{epsfig}

\def\beq{\begin{equation}}
\def\eeq{\end{equation}}
\def\bea{\begin{eqnarray}}
\def\eea{\end{eqnarray}}

\begin{document}
\input epsf.tex
 
\title{The Search for Massive Black Hole Binaries with LISA.}
\author{Neil J. Cornish and Edward K. Porter} 
\address{Department of Physics, Montana State University, Bozeman, 59717, MT, USA.}
\vspace{1cm}
\begin{abstract}
\noindent In this work we focus on the search and detection of Massive black hole binary (MBHB) systems, including systems at high redshift.  As well as expanding on previous works where we used a variant of Markov Chain Monte Carlo (MCMC), called Metropolis-Hastings Monte Carlo, with simulated annealing, we introduce a new search method based on frequency annealing which leads to a more rapid and robust detection.  We compare the two search methods on systems where we do and do not see the merger of the black holes.  In the non-merger case, we also examine the posterior distribution exploration using a 7-D MCMC algorithm.  We demonstrate that this method is effective in dealing with the high correlations between parameters, has a higher acceptance rate than previously proposed methods and produces posterior distribution functions that are close to the prediction from the Fisher Information matrix.  Finally, after carrying out searches where there is only one binary in the data stream, we examine the case where two black hole binaries are present in the same data stream.  We demonstrate that our search algorithm can accurately recover both binaries, and more importantly showing that we can safely extract the MBHB sources without contaminating the rest of the data stream.  \\ \linebreak
\end{abstract}

\maketitle

\section{Introduction}
\subsection{Background.}
Massive black hole binaries (MBHB) are expected to be one of the strongest candidate sources for LISA, the Laser Interferometer Space Antenna, a joint ESA-NASA mission that will search for gravitational waves (GW) is the frequency bandwidth $10^{-5}\leq f/Hz \leq 1$~\cite{LISA}.  There is growing observational evidence for the existence of Massive black holes (MBH) at the center of most galaxies.  Galactic formation models, such as Hierarchical structure formation, suggest that modern day MBHs are the direct result of the coalescence of smaller "seed" black holes of $\sim 10^{5}M_{\odot}$ throughout cosmic history~\cite{seedgal}.  Observations also suggest direct evidence of a MBHB in the radio galaxy 0402+379~\cite{RTZPPR}, at the center of the Quasar 3C 345~\cite{LR}.  While there are other possible detections, it is the system 0402+379 that seems to give the most compelling evidence of a MBHB.  It is suggested that at the center of this system there are two MBHs with a combined total mass of $\sim 10^{8}M_{\odot}$ and an orbital separation of 7.3 pc. 

The detection of MBHBs by LISA is important for two main reasons.  Firstly, it will allow us to carry out a test of gravity in the highly nonlinear strong-field regime~\cite{Ryan, colhugh, bercord}.  Secondly, it will allow us, in conjunction with other astronomical methods, to investigate such things as galaxy interactions and mergers out to very high redshift $(z\geq 10)$.  These systems are also very attractive sources for LISA.  Due to their high masses, even as these systems approach merger and ringdown, they occur at frequencies which are blocked to the ground based detectors because of such things as tectonic motion below about 10 Hz and more importantly, time varying gravitational potentials such as weather systems etc.  Also, unlike galactic binaries and Extreme Mass Ratio Inspirals (EMRIs), the MBHBs are very clean sources with measureable signal to noise ratios (SNR) of order $\sim 100 - 1000$'s.  While the actual event rate for MBHBs is still a subject of contention~\cite{Berti1, sesena}, it is unlikely to be high enough to lead to confusion noise between sources, something which is very important in the search for galactic binaries and to a lesser extent in the search for EMRIs.

While a lot of work has been done on source modelling, especially for EMRI sources, there has not been much research done into developing viable detection algorithms for the LISA data analysis problem.  One option would be to adopt the current ground based inspiral methods and use a template bank to search for each source.  However, due to the high cost of templates involved for a blind template bank searching for a singe source ($10^{7}$ for galactic binaries~\cite{en1}, $10^{13}$ for non-spinning MBHBs~\cite{en2} and $10^{40}$ for EMRIs~\cite{gairetal}) other methods are needed.  While there are currently other methods being developed, such as genetic algorithms~\cite{ccr}, iterative methods~\cite{cl,rch}, time-frequency methods~\cite{gw} and tomographic reconstruction~\cite{mn}, we present here a variant of the Markov Chain Monte Carlo method (MCMC)~\cite{mcmc1, mcmc2} called Metropolis-Hastings Monte Carlo (MHMC).  The MCMC algorithm is a useful approach to searching through large parameter spaces, performing model comparisons, estimating instrument noise and providing error estimates.  It has already been successfully been used in other astronomical fields such as cosmological parameter estimation with the WMAP data~\cite{wmap}.  In terms of GWs, the MCMC technique has already been applied to ground-based analysis~\cite{ch1, ch2, ch3}, a LISA toy-model problem~\cite{andrieu, umstat} and galactic binaries~\cite{cc1,cc2}. For MBHBs both MCMC and MHMC techniques have been successfully applied~\cite{en2, en3, vecchio1}. 

In this work, we will focus on the detection of gravitational waves from inspiralling Schwarzschild BHs in the Low Frequency Approximation.  In the future we intend extending our work to include extra complications such as higher harmonic corrections to the amplitude, introducing spins and more complicated detector response functions to take into account lower mass systems.

\subsection{The LISA detector and response at the detector to GWs.}
LISA is composed of a constellation of three drag-free spacecraft forming an equilateral triangle with each side of length $L = 5\times 10^{6}$ km.  The center of mass of the detector follows a heliocentric orbit $20^{o}$ behind the Earth.  The detector triangle is tilted by $60^{o}$ with respect to the ecliptic.  Due to orbits of each individual spacecraft being on different planes, the detector triangle rotates about itself with a period of 1 year.  The detector itself is an all-sky detector.  The LISA configuration allows us to use the three detector arms as a pair of two arm detectors.  The motion of the detector around the Sun introduces a periodic Doppler shift whose magnitude and phase are dependent on the angular position of the source in the sky.

While the full response from the detector involves issues such as fluctuations of arm-length, pointing ahead and signal cancellation once the wavelength of the GW becomes comparable to the armlength of LISA, we can make an excellent 1st approximation called the Low Frequency Approximation (LFA)~\cite{cutler98}.  By assuming that there are no fluctuations in arm length, ignoring the problem of pointing ahead, ignoring signal cancellation due to the LISA transfer functions and assuming we can evaluate all three spacecraft at the same time, we can safely work in the LFA.  Also, as long as the frequency of the MBH is less than the transfer frequency of the detector, i.e. $f \ll f_{*}\sim 10^{-2}$ Hz, the LFA is a good approximation to the total response from the detector.  In fact it was shown that the LFA is a good approximation to the full detector response to a frequency of $\sim 3$ mHz~\cite{cornishrubbo}   In this approximation the beam pattern functions are essentially a quadrupole antennae.

Working in the limit $f\ll f_{*}$ and $f/\dot{f} \ll L$, we can express the strain at the detector to an incoming GW with polarizations $h_{+,\times}(t)$  as
\begin{equation}
h(t) = h_{+}(\xi(t))F^{+}+h_{\times}(\xi(t))F^{\times}, 
\end{equation}
where the phase shifted time parameter is 
\begin{equation}
\xi(t) = t - R_{\oplus}\sin\theta\cos\left(\alpha(t) - \phi\right).
\end{equation}
Here, $R_{\oplus} = 1 AU \approx$ 500 secs is the radial distance to the detector guiding center, $\left(\theta,\phi\right)$ are the position angles of the source in the sky, $\alpha(t)=2\pi f_{m}t + \kappa$, $f_{m}=1/year$ is the LISA modulation frequency and $\kappa$ gives the initial ecliptic longitude of the guiding center.  The beam pattern functions are defined by
\beq
F^{+}(t) = \frac{1}{2}\left[\cos(2\psi)D^{+}(t;\theta, \phi, \lambda) - \sin(2\psi)D^{\times}(t;\theta, \phi, \lambda)\right],
\eeq
\beq
F^{\times}(t) = \frac{1}{2}\left[\sin(2\psi)D^{+}(t;\theta, \phi, \lambda) + \cos(2\psi)D^{\times}(t;\theta, \phi, \lambda)\right].
\eeq
The quantity $\psi$ is the polarization angle of the wave.  Formally, if ${\bf \hat{L}}$ is the direction of the binary's orbital angular momentum and ${\bf\hat{n}}$ is the direction from the observer to the source (such that the GWs propagate in the ${-\bf\hat{n}}$ direction), then $\psi$ fixes the orientation of the component of ${\bf\hat{L}}$ perpendicular to ${\bf\hat{n}}$.  The time dependent quantities $D_{+,\times}(t)$ are given in the LFA by~\cite{cornishrubbo} 
\bea
\fl D^{+}(t) = \frac{\sqrt{3}}{64}\left[\frac{}{}-36\sin^{2}(\theta)\sin(2\alpha(t)-2\lambda)+(3+\cos(2\theta)) \right.\\ \nonumber\fl \left(\frac{}{}\cos(2\phi)\left\{\frac{}{}9\sin(2\lambda)-\sin(4\alpha(t)-2\lambda)\right\} \frac{}{}+\sin(2\phi)\left\{\frac{}{}\cos\left(4\alpha(t)-2\lambda\right)-9\cos(2\lambda) \right\}\frac{}{}\right)\\ \nonumber  \left.-4\sqrt{3}\sin(2\theta)\left(\frac{}{}\sin(3\alpha(t)-2\lambda-\phi)-3\sin(\alpha(t)-2\lambda+\phi)\right)\right]
\eea
\bea
\fl D^{\times}(t) = \frac{1}{16}\left[\frac{}{}\sqrt{3}\cos(\theta)\left(\frac{}{}9\cos(2\lambda-2\phi)-\cos(4\alpha(t)-2\lambda-2\phi) \right) \right. \\ \nonumber \left. -6\sin(\theta)\left(\frac{}{} \cos(3\alpha(t)-2\lambda-\phi)+3\cos(\alpha(t)-2\lambda+\phi) \right) \right],
\eea
where $\lambda = 0$ and $\pi/4$ give the orientation of the two detectors.  We can see from the above equations that we do not measure the polarizations individually at each detector, but a combination of the polarizations weighted by the beam pattern functions.

\subsection{Organization of the paper.}
The aim of this paper is to extend on previous works on the search for GWs from a MBHB~\cite{en2, en3, vecchio1} using the LISA detector.  In these previous works it was shown that it was possible to carry out not only a search for these sources, but also to map out the posterior distributions  for the errors in the parameter estimation of the source.

The organization of the rest of the paper is as follows : In Section~\ref{sec:gw} we define the gravitational waveform describing the inspiral phase of the two MBHs.  We outline the division of the parameters into those extrinsic and intrinsic to the source.  We then go on to define the restricted post-Newtonian (PN) approximation and finally state the post-Newtonian equations describing both the phase and frequency evolution of the waveform.

In Section~\ref{sec:da} we outline the theory behind estimating both the source parameters and the errors in such an estimation.  We briefly outline the geometric nature of the problem, before going on to outline the main tools behind our analysis.  This will entail a definition of the F-Statistic method and a discussion on the confusion noise from the galactic foreground.
 
In Section~\ref{sec:method} we present a discussion of Metropolis-Hastings sampling using simulated annealing.  We will also present a new alternative method which we describe as frequency annealing.  

Section~\ref{sec:results1} contains a presentation of the results from our analysis as we search for single sources in the data stream.  Because of the brightness of MBHBs, it is actually quite difficult to define an astrophysically reasonable low SNR source in the case where we observe the coalescence of the MBHs.  While it not hard to have galactic binaries with an SNR of $\sim5$,  to define MBHBs with SNR of $\sim10$ we have to assume that the time to coalescence is much greater than the time of observation.  In a previous analysis~\cite{en2}, we have found that it is quite trivial to find sources with SNRs of $\geq$ 100.

It is because of this that we have chosen sources with relatively low SNR, covering various mass ranges from almost equal to a mass ratio of about 10, with distances up to $z = 10$ and with total mass ranges that show that it should be possible for LISA to see the coalescence of "seed" galaxies at cosmological distances.  To make the problem more realistic, we also introduce a galactic foreground of approximately 26 million individually modelled galactic binary sources.  This galactic foreground is generated using a modern population synthesis code~\cite{NYZ} and a fast detector response code \cite{NT, MLDC2}.  At the end of this section we discuss the Markov Chain Monte Carlo method we use to analyze the posterior distributions.  One of the important parts of this section is a discussion on how to compensate for the large correlations between the phase at coalescence and the other parameters when we explore the posteriors in the case where we do not see coalescence.

Section~\ref{sec:posteriors} contains an analysis of the exploration of the posterior distribution functions for each of the parameters.  We concentrate on the case where we do not see the coalescence of the waveform as this is the more challenging situation.  In the case where coalescence is seen, the phase at coalescence is sufficiently resolved that it minimizes the correlations between the parameters.  We have addressed this issue in a previous publication~\cite{en3}.  When we do not see the coalescence of the waveform, the phase at coalescence is virtually undetermined.  We show from looking at a slice through the Likelihood surface shows that the error ellipsoid is very long and narrowly peaked, like the blade of a knife.  This is a problem for any MCMC exploration as most proposals will be rejected due to the fact that it is so difficult to stay on the knife edge.  We propose a scheme where we maximize over the phase at coalescence and the luminosity distance and conduct our exploration in 7 dimensions only.

In Section~\ref{sec:msmbh} we examine the search when there are two MBHBs present in the data stream.  We demonstrate that because of the small overlap between two MBHB sources that we can pull out each of the MBHBs individually with virtually no contamination to the rest of the data.  Our algorithm at present uses an iterative search for the MBHBs.

\section{The Gravitational Waveform}\label{sec:gw}
The gravitational waveform from inspiralling bodies is composed of an inspiral, merger and ringdown phase.  While we believe we have good knowledge of the inspiral and ringdown phases, the merger phase is less well understood, though there has been great recent progress~\cite{numerical}.  Due to the recent progress in Numerical Relativity, it is hoped that some time in the near future we will be able to model the entire waveform.  For now, we focus only on the initial inspiral phase. Such an inspiral waveform is referred to as a chirp waveform due to the increase of amplitude and frequency caused by a slow decay in the orbit of the two component bodies.  This decay is caused by the loss in energy and angular momentum from the system in the form of GWs.  In this analysis, we focus on MBHBs where the component bodies are Schwarzschild MBHs.  In this case the waveform is described by the nine parameter set $\vec{x}=\{\ln(M_{c}),\ln(\mu),\theta, \phi, \ln(t_{c}), \iota, \varphi_{c}, \ln(D_{L}), \psi\}$, where $M_{c}$ is the chirp mass, $\mu$ is the reduced-mass, $(\theta,\phi)$ are the sky location of the source, $t_{c}$ is the time-to-coalescence, $\iota$ is the inclination of the orbit of the binary, $\varphi_{c}$ is the phase of the GW at coalescence, $D_{L}$ is the luminosity distance and $\psi$ is the polarization of the GW.  We will describe the parameter subset $\{D_{L}, \iota, \varphi_{c}, \psi\}$ as being extrinsic parameters, while all the rest will be described as being intrinsic, i.e. parameters that describe the dynamics of the binary.  We should also mention that $(\theta,\phi)$, which would normally be classed as being extrinsic, are classed as being intrinsic due to the fact that they are tied to the motion of LISA through the beam pattern functions.

The GWs are defined as a superposition of harmonics at multiples of the orbital phase.  The two harmonics of the wave can be written as
\begin{equation}
h_{+, \times}(t) = Re\left\{\sum_{m}H_{+,\times}^{(m)}(t)e^{im\Phi_{orb}}\right\},
\end{equation}
where $m$ is the harmonic index, $\Phi_{orb}$ is the orbital phase and $H_{+,\times}^{(m)}$ are the amplitude corrections associated with each harmonic.  The strongest harmonic of the wave is the one given by the quadrupole moment of the source, corresponding to $m=2$.  In the restricted post-Newtonian approximation we neglect all other harmonics besides the quadrupole term in the amplitude, but expand the corrections in the phase of the wave up to n-PN order, i.e. $\sim(v/c)^{2n}$.  We should point out that this approximation neglects the phase modulations introduced by the other amplitude corrections.  In a future publication we will examine the consequences of searching for a signal with the full amplitude corrections using the restricted PN waveforms.  For this study, we include corrections to the phase up to 2-PN order.

In the restricted post-Newtonian approximation, the GW polarizations are given by~\cite{biww}
\begin{eqnarray}
h_{+}& =& \frac{2Gm\eta}{c^{2}D_{L}}\left(1+\cos^{2}\iota\right)x\cos(\Phi),\\ \nonumber \\
h_{\times} &= &-\frac{4Gm\eta}{c^{2}D_{L}}\cos\iota\,x\sin(\Phi).
\end{eqnarray}
Here $m=m_{1}+m_{2}$ is the total mass of the binary, $\eta = m_{1}m_{2}/m^{2}$ is the reduced mass ratio, $G$ is Newton's constant and $c$ is the speed of light.  The inclination of the orbit of the binary system is formally defined as $\cos\iota=\bf\hat{L}\cdot\hat{n}$.  The invariant PN velocity parameter is defined by $x = \left(Gm\omega / c^{3}\right)^{2/3}$, where 
\begin{eqnarray}\label{eqn:freq}
\omega(t)&=&\frac{c^{3}}{8Gm}\left[\Theta^{-3/8}+\left(\frac{743}{2688}+\frac{11}{32}\eta\right)\Theta^{-5/8}-\frac{3\pi}{10}\Theta^{-3/4}\right.\nonumber\\ &+&\left.\left(\frac{1855099}{14450688}+\frac{56975}{258048}\eta+\frac{371}{2048}\eta^{2}\right)\Theta^{-7/8}\right]\nonumber\\
\end{eqnarray}
is the 2 PN order orbital frequency for a circular orbit formally defined as $\omega=d\Phi_{orb}/dt$, and $\Phi =\varphi_{c}-\varphi(t) = 2\Phi_{orb}$ is the gravitational wave phase which is defined as
\begin{eqnarray}\label{eqn:phase}
\Phi(t) &=& \varphi_{c}-\frac{2}{\eta}\left[\Theta^{5/8}+\left(\frac{3715}{8064}+\frac{55}{96}\eta\right)\Theta^{3/8}-\frac{3\pi}{4}\Theta^{1/4}\right.\nonumber\\ &+&\left.\left(\frac{9275495}{14450688}+\frac{284875}{258048}\eta+\frac{1855}{2048}\eta^{2}\right)\Theta^{1/8}\right].\nonumber\\
\end{eqnarray}
The time dependent quantity $\Theta(t;t_{c})$ is related to the time to coalescence of the wave, $t_{c}$, by
\begin{equation}
\Theta(t;t_{c}) = \frac{c^{3}\eta}{5Gm}\left(t_{c}-t\right).
\end{equation}
As they are used interchangeably in our analysis, the relationships between $(m,\eta)$ and $(M_{c}, \mu)$ are given by $M_{c}=m\eta^{3/5}$ and $\mu = m\eta$.  In Table~\ref{tab:paramsandpriors} we present the parameter values and the prior ranges for both MBHBs that we use in this analysis.  We should point out that the individual masses quoted are the rest-frame masses, while the maximum GW frequency reached and the priors for the total mass are given in redshifted values.  The priors were chosen to encompass an astrophysically interesting range, and also not tight enough as to make sure that the chain would definitely find the sources.

\begin{table}[t]
\begin{center}
\begin{tabular}{|c|c|c|}\hline\hline
&  &  \\
 & Case A & Case B \\ &  &  \\ \hline &  &  \\
$m_{1}/M_{\odot}$ & $\,\,\,\,\,\,\,\,5\times10^{5}\,\,\,\,\,\,\,\,$ & $\,\,\,\,\,\,\,\,1\times10^{6}\,\,\,\,\,\,\,\,$ \\
$m_{2}/M_{\odot}$ & $3\times10^{5}$ & $2\times10^{5}$ \\
$\theta / rad$ & 0.6842 & 1.6398\\
$\phi / rad$  & 2.5791 & 4.1226\\
$t_{c}/yr$ & 0.458333 & 1.04\\
$\iota / rad$ & 0.9273 & 0.7647\\
$\psi / rad$ & 1.4392 & 1.373\\
$z$ & 10 & 1.5 \\
$D_{L} / Gpc$ & 106.28 & 11.01\\
$\varphi_{c} / rad$ & 0.3291 & 0.9849\\
$f_{gw}^{max}(z) / mHz$ & $0.3965$ & $0.14584$\\
$m_{z}^{min} / M_{\odot}$ & $5\times10^{6}$ & $1\times10^{6}$\\
$m_{z}^{max} / M_{\odot}$ & $5\times10^{7}$ & $1\times10^{7}$\\
$(m_{1}/m_{2})_{min}$ & 5 & 15\\
$(m_{1}/m_{2})_{max}$ & 1 & 5\\
$t_{c}^{min} / yr$ & 0.25 & 0.75\\ 
$t_{c}^{max} / yr$ & 0.75 & 1.25\\
$SNR$ & 40.989 & 97.39\\  
&  &    \\ 
\hline\hline
\end{tabular}
\end{center}
\caption{The parameter values and priors for the two test cases studied.  In the above table, both the maximum gravitational wave frequency and total mass priors quoted are redshifted values.}
\label{tab:paramsandpriors}
\end{table}

While we initially define our sources in terms of redshift $z$, it is the luminosity distance $D_{L}$ that enters into the waveform equations.  Using the WMAP values of $(\Omega_{R}, \Omega_{M}, \Omega_{\Lambda}) = (4.9\times10^{-5}, 0.27, 0.73)$ and a Hubble's constant of $H_{0}$=71 km/s/Mpc~\cite{wmap}, the relation between redshift, $z$, and luminosity distance, $D_{L}$ is given by
\begin{equation}
D_{L} = \frac{c(1+z)}{H_{0}}\int_{0}^{z}\,dz'\left[\Omega_{R}\left(1+z'\right)^{4}+\Omega_{M}\left(1+z'\right)^{3}+\Omega_{\Lambda}\right]^{-1/2}.  
\end{equation}
   
For two Schwarzschild black holes, Equations~(\ref{eqn:freq}) and (\ref{eqn:phase}) governing the evolution of the frequency and phase break down even before we reach the last stable circular orbit (LSO) at $R=6M$.  Because of this, if we are examining a case where the source coalesces within the observation period,  we terminate our waveforms at $R=7M$.   Due to the fact that we are using numerical Fourier transforms in our analysis, if we just truncate the time domain waveform, we get spectral leakage and intense ringing in the Fourier domain.  It is also a requirement of the Fourier routines that we use that each waveform array is an integer power of 2.  We therefore need to truncate our waveform arrays smoothly and pad the rest of the array with zeros.  In order to do this, we introduce the following hyperbolic taper function
\begin{equation}
{\mathcal F}(x, x_{max})=\frac{1}{2}[1-\tanh(k\left\{x-x_{max}\right\})].
\end{equation}
with which we multiply the amplitude of the wave.  In the case where we see coalescence, we set $x_{max}=1/7$.  In the case where we do not see coalescence, we use the following methodology.  Solving a non-linear equation in $\Theta$ of the form $g(\theta)=1/7$, we find $\Theta_{min}(t)$, and from this calculate the maximum frequency, and hence the maximum velocity reached during the time of observation.  We then set this value of $x_{max}$ in the taper, and once again pad the waveform arrays with zeros after this velocity has been reached.  The steepness parameter $k$ in the taper is chosen to be sufficiently large ($\geq 100$) that it cuts off the time domain waveform smoothly, while reducing the ringing in the frequency domain.

\section{Measuring and estimating parameter errors with LISA.}\label{sec:da}
\subsection{Background theory.}
Most of the current methods of parameter measurement and error estimation are based on a geometric model of signal analysis~\cite{Helst, Owen, porter}.  In this treatment, the waveforms are treated as vectors in a Hilbert space.  This vector space has a natural scalar product
\begin{equation}\label{eqn:scalarprod}
\left<h\left|s\right.\right> =2\int_{0}^{\infty}\frac{df}{S_{n}(f)}\,\left[ \tilde{h}(f)\tilde{s}^{*}(f) +  \tilde{h}^{*}(f)\tilde{s}(f) \right].
\label{eq:scalarprod}
\end{equation}
and vector norm $|h| = \left<h\left|h\right.\right>^{1/2}$, where
\begin{equation}
\tilde{h}(f) = \int_{-\infty}^{\infty}\, dt\, h(t)e^{2\pi\imath ft}
\end{equation}
is the Fourier transform of the time domain waveform $h(t)$.  The quantity $S_{n}(f)$ is the one-sided noise spectral density of the detector and will be defined at a later stage.

As previously stated, we can think of the three arms of the LISA interferometer as being two separate $90^{o}$ interferometers.  In this case, the signal in each of the detectors is given by
\begin{equation}
s_{i}(t) = h_{i}(t)+n_{i}(t),
\end{equation}
where $i=I, II$ label each detector.  We assume that the noise $n_{i}(t)$ is stationary, Gaussian, uncorrelated in each detector and characterized by the noise spectral density $S_{n}(f)$.  Using the scalar product defined in Equation~(\ref{eqn:scalarprod}), we define the SNR in each detector as 
\begin{equation}
\rho_{i} = \frac{\left<h\left|s_{i}\right.\right>}{\sqrt{\left<h\left|h\right.\right>}}.
\end{equation}
Closely related to the SNR is the normalized overlap between two templates defined by 
\begin{equation}
{\mathcal O} = \frac{\left<h_{1}\left|h_{2}\right.\right>}{\sqrt{\left<h_{1}\left|h_{1}\right.\right>\left<h_{2}\left|h_{2}\right.\right>}}.
\end{equation}
Now, given some signal $s(t)$, the likelihood that the true parameter values are given by some parameter vector $\vec{x}$ is described by 
\begin{equation}\label{eqn:likelihood}
{\mathcal L}\left(\vec{x}\right) = C\,e^{-\left<s-h\left(\vec{x}\right)|s-h\left(\vec{x}\right)\right>/2},  
\end{equation}
where $C$ is a normalization constant.  To achieve the Maximum Likelihood, we search for a parameter set that minimizes the exponent in the above equation.  For most MBHBs, the SNR will be high.  In this high SNR limit, the errors in the parameter estimation will have a Gaussian probability distribution given by
\begin{equation}
p\left(\Delta\vec{x}\right)=\sqrt{\frac{\Gamma}{2\pi}}e^{-\frac{1}{2}\Gamma_{\mu\nu}\Delta x^{\mu}\Delta x^{\nu}},
\end{equation}
where $\Gamma_{\mu\nu}$ is the Fisher information matrix (FIM)
\begin{equation}
\Gamma_{\mu\nu} = \left<\frac{\partial h}{\partial x^{\mu}}\left|\frac{\partial h}{\partial x^{\nu}}\right.\right>,
\end{equation}
and $\Gamma = det\left(\Gamma_{\mu\nu}\right)$.  When we use both LISA detectors the total SNR is given by
\begin{equation}
\rho = \rho^{I} + \rho^{II},
\end{equation}
while the total FIM is given by
\begin{equation}
\Gamma_{\mu\nu} = \Gamma_{\mu\nu}^{I} + \Gamma_{\mu\nu}^{II}.
\end{equation}
Once we have the total FIM, we numerically invert to get the variance-covariance matrix
\begin{equation}
C^{\mu\nu} = \Gamma_{\mu\nu}^{-1}.
\end{equation}
The diagonal elements of $C^{\mu\nu}$ give a $1\sigma^{2}$ estimate of the error in the parameter estimation, i.e.
\begin{equation}
\Delta x^{\mu} = \sqrt{C^{\mu\mu}},
\end{equation}
while the off-diagonal elements give the correlations between the various parameters
\begin{equation}
c^{\mu\nu} = \frac{C^{\mu\nu}}{\sqrt{C^{\mu\mu}C^{\nu\nu}}}\,\,\,\,\,\,\,\,\,\,\,\,-1\leq c^{\mu\nu}\leq 1.
\end{equation}
Theoretically, in the large SNR limit, $C^{\mu\nu}$ is the Cramer-Rao bound.  The problem is that it is very difficult so say what exactly constitutes the large SNR limit.  For low SNRs, the assumption that the errors have a Gaussian probability distribution breaks down.

\begin{figure}[t]
\begin{center}
\epsfig{file=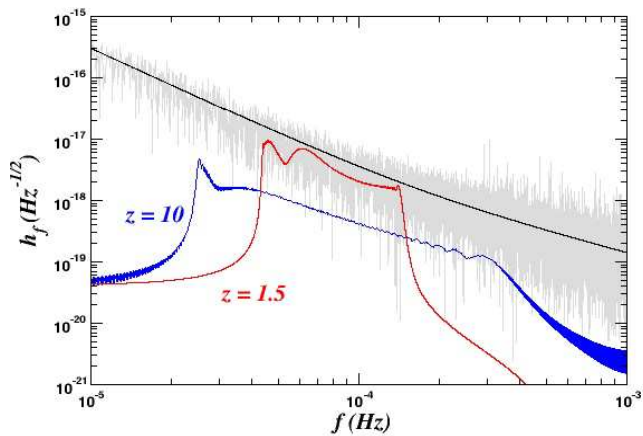, width=3.5in, height=2.75in}
\end{center}
\caption{A plot of the power spectra of our two test sources.  The spectra of the signals is plotted against the combined instrument noise and galactic foreground.  The solid line running through the noise is the rms noise for the combined noise sources.}
\label{fig:sigspectra}
\end{figure}

\subsection{The noise power spectral density.}
We can see from the above equations that the scalar products and calculation of the FIM are all weighted with the one-sided power spectral density $S_{n}(f)$. The noise spectral density in the detector can be written in terms of the autocorrelation of the noise
\begin{equation}
\left<\tilde{n}(f)\tilde{n}^{*}(f')\right> = \frac{T_{obs}}{2}\delta_{ff'}S_{n}(f),
\end{equation}
where $T_{obs}$ is the time period of observation and the angular brackets denote an expectation value.  The total noise power spectral density in our analysis can be broken into two parts : instrumental and galactic.  To model the instrumental noise we use the standard one-sided noise spectral density for the LISA detector given by~\cite{cornish}
\begin{equation}
S_{n}^{ins}(f)=\frac{1}{4L^{2}}\left[ S_{n}^{pos}(f)+2\left(1+\cos^{2}\left(\frac{f}{f_{*}}\right) \right)\frac{S_{n}^{acc}(f)}{(2\pi f)^{4}}\right] , 
\end{equation}
where $L=5\times10^{6}$ km is the arm-length for LISA,  $S_{n}^{pos}(f) = 4\times10^{-22}\,m^{2}/Hz$ and $S_{n}^{acc}(f) = 9\times10^{-30}\,m^{2}/s^{4}/Hz$ are the position and acceleration noise respectively.  The quantity $f_{*}=1/(2\pi L)$ is the mean transfer frequency for the LISA arm.  Using this above formula, we generate instrumental noise from a random Gaussian distribution. 

As well as the instrumental noise, the other main source of noise comes from the galactic population of binaries.  To simulate the galactic foreground we generate the response to 26 million individually modelled galactic binaries and superimpose the responses~\cite{MLDC2}.  While some of these binaries are bright enough that they will be individually resolved, the vast majority will not be.  The millions of unresolvable galactic binaries which constitute the galactic foreground produce a noise source which is commonly referred to as confusion noise.  To model the galactic or confusion noise we use the following confusion noise estimate derived from a Nelemans, Yungelson, Zwart (NYZ) galactic foreground model~\cite{NYZ, TRC}
\begin{equation}
S_{n}^{conf}(f) = \left\{ \begin{array}{ll} 10^{-44.62}f^{-2.3} & 10^{-4} < f\leq 10^{-3} \\ \\ 10^{-50.92}f^{-4.4} & 10^{-3} < f\leq 10^{-2.7}\\ \\ 10^{-62.8}f^{-8.8} &  10^{-2.7} < f\leq 10^{-2.4}\\ \\ 10^{-89.68}f^{-20} &  10^{-2.4} < f\leq 10^{-2}  \end{array}\right.,
\end{equation}
where the confusion noise has units of $m^{2}Hz^{-1}$.  In our search analysis, we weight our scalar products with the total power spectral density
\begin{equation}
S_{n}(f) = S_{n}^{ins}(f)+ S_{n}^{conf}(f).
\end{equation}
In Figure~\ref{fig:sigspectra} we plot the power spectra of both sources against the instrument noise and the galactic foreground of approximately 26 million sources.  The solid line in the figure is the rms noise of the combined noise sources.  We should mention here, that once we introduce the galactic foreground, we automatically lose the assumption that the noise is stationary and Gaussian.  We will see later in the paper how this effects the statistics of the recovered parameter values.

\subsection{The generalized F-Statistic.}
One of the advantages of dividing the parameter set into extrinsic and intrinsic, is that we can project onto the orthogonal subspace defined by the intrinsic parameters to reduce the dimensionality of the search.  Instead of having to search over the full 9-d search space, we only need to search the subset of intrinsic parameters $\{\ln(M_{c}), \ln(\mu), \ln(t_{c}), \theta,\phi\}$.  We can use the F-statistic~\cite{jks} to find the optimal values of the four extrinsic parameters $\{\iota, \psi, D_{L}, \varphi_{c}\}$.  This is done as follows.  We first write the strain of the gravitational wave in the form
\begin{equation}
h(t) = \sum_{i=1}^{4}a_{i}\left(\iota, \psi, D_{L}, \varphi_{c}\right)A^{i}\left(t;M_{c}, \mu, t_{c}, \theta,\phi\right),
\label{eqn:strainFS}
\end{equation}
where
\begin{eqnarray}
a_{1}& =& \Lambda\left[\left(1 + \cos^{2}\iota\right)\cos2\psi\cos\varphi_{c}-2\cos\iota\sin2\psi\sin\varphi_{c} \right]\nonumber\\
a_{2} &= &-\Lambda\left[\left(1 +\cos^{2}\iota\right)\sin2\psi\cos\varphi_{c}+2\cos\iota\cos2\psi\sin\varphi_{c} \right]\nonumber\\
a_{3} &= &\Lambda\left[\left(1 +\cos^{2}\iota\right)\cos2\psi\sin\varphi_{c}+2\cos\iota\sin2\psi\cos\varphi_{c} \right]\nonumber\\
a_{4} &=& -\Lambda\left[\left(1 +\cos^{2}\iota\right)\sin2\psi\sin\varphi_{c}-2\cos\iota\cos2\psi\cos\varphi_{c} \right],\nonumber\\
\end{eqnarray}
and
\begin{eqnarray}
A^{1} &=& m_{o}\eta\,x(t)\,D^{+}\cos(\varphi)\nonumber\\
A^{2} &=& m_{o}\eta\,x(t)\,D^{\times}\cos(\varphi)\nonumber\\
A^{3} &=& m_{o}\eta\,x(t)\,D^{+}\sin(\varphi)\nonumber\\
A^{4} &=& m_{o}\eta\,x(t)\,D^{\times}\sin(\varphi).
\end{eqnarray}
Here $\Lambda = c / D_{L}$ and $m_{o} = Gm/c^{3}$ is the total mass in seconds.  We can see that the time-independent $a_{i}$ quantities give us a series of four equations in four unknowns.  By defining four constants $N^{i}=\left<s \left| A^{i}\right>\right.$, where $s(t)=h(t)+n(t)$ is the signal, we can find a solution for the $a_{i}$'s in the form
\begin{equation}
a_{i} = M_{ij}N^{j},
\label{eqn:ais}
\end{equation}
where the M-Matrix is defined by
\begin{equation}
M_{ij} = \left(M^{ij}\right)^{-1} = \left<A^{i} \left| A^{j}\right>^{-1}\right. .
\end{equation}
If we now substitute the above equation into the expression for the reduced log-likelihood, i.e.
\begin{equation}
\ln \,{\mathcal L}\left(\vec{x}\right) = \left<s \left| h\left(\vec{x}\right)\right>\right. - \frac{1}{2}\left<h\left(\vec{x}\right) \left| h\left(\vec{x}\right)\right>\right. ,
\end{equation}
we obtain the F-statistic
\begin{equation}
{\mathcal F} = \frac{1}{2}M_{ij}N^{i}N^{j},
\end{equation}
which automatically maximizes the log-likelihood over the extrinsic parameters and reduces the search space to the sub-space of intrinsic parameters.  Once we have the numerical $a_{i}$'s, we can set about finding the maximized extrinsic parameters.  Defining the quantities
\begin{equation}
A_{+}=\sqrt{\left(a_{1}-a_{4}\right)^{2} + \left(a_{2}+a_{3}\right)^{2}} + \sqrt{\left(a_{1}+a_{4}\right)^{2} + \left(a_{2}-a_{3}\right)^{2}},
\end{equation}
\begin{equation}
A_{\times}=\sqrt{\left(a_{1}-a_{4}\right)^{2} + \left(a_{2}+a_{3}\right)^{2}} - \sqrt{\left(a_{1}+a_{4}\right)^{2} + \left(a_{2}-a_{3}\right)^{2}},
\end{equation}
and
\begin{equation}
A = A_{+} + \sqrt{A_{+}^{2} + A_{\times}^{2}},
\end{equation}
we find solutions for the four extrinsic parameters in the form
\begin{eqnarray}
\iota & = & \arccos\left(\frac{-A_{\times}}{A}\right),\\
\psi & = & \frac{1}{2}\arctan\left(\frac{-\left(A_{+}a_{4}+A_{\times}a_{1}\right)}{-\left(A_{\times}a_{2}-A_{+}a_{3}\right)}\right),\\
D_{L} &=& \frac{4c}{A},\\
\varphi_{c}& = &\arctan\left(\frac{-c\left(A_{+}a_{4}+A_{\times}a_{1}\right)}{-c\left(A_{+}a_{2}-A_{\times}a_{3}\right)}\right),
\end{eqnarray}
where $c = sgn(\sin(2\psi))$.  We should note that the F-Statistic is related to the SNR by
\begin{equation}
{\mathcal F} \approx \left< \ln {\mathcal L}\right> = \frac{\rho^{2}+D}{2},
\end{equation}
where in this case the angular brackets again denote the expectation value and $D$ is the dimensionality of the search space.

\section{Metropolis-Hastings sampling with simulated and frequency annealing.}\label{sec:method}
The Metropolis-Hastings sampling method is a variant on the Markov Chain Monte Carlo method (MCMC), and works as follows : starting with the signal $s(t)$ and some initial template $h(t)$, we choose a starting point randomly in the parameter space $\vec{x}$.  We then draw from a proposal distribution and propose a jump to another point in the space $\vec{y}$.  In order to compare both points, we evaluate the Metropolis-Hastings ratio
\begin{equation}
H = \frac{\pi(\vec{y})p(s|\vec{y})q(\vec{x}|\vec{y})}{\pi(\vec{x})p(s|\vec{x})q(\vec{y}|\vec{x})}.
\end{equation}
Here $\pi(\vec{x})$ are the priors of the parameters, $p(s|\vec{x})$ is the likelihood, and $q(\vec{x}|\vec{y})$ is the proposal distribution.  This jump is then accepted with probability $\alpha = min(1,H)$, otherwise the chain stays at $\vec{x}$. 

The above steps are the basis of most MCMC methods.  However, as a search algorithm on its own, the MCMC method is inefficient and would take too long to find the source we are looking for.  It also suffers from other limitations.  If the source is particularly bright and we start too far away from the true parameter values, it is very easy for the chain to get stuck on a secondary or tertiary maximum.  Once the chain gets stuck, it is very difficult for it to move off the local maximum.  In order to actually convert the MCMC method into a bona fide search algorithm, we need to use simulated annealing and a variety of proposal distributions that allow a range of different size jumps in the parameter space.  We also need to include other non-Markovian methods that we will describe below.  It is this combination of Metropolis-Hastings and non-Markovian methods that we term Metropolis-Hastings Monte Carlo (MHMC).  First of all we will describe the various proposal distributions used in our algorithm, followed by $t_{c}$ maximization and simulated annealing.

In order to make small jumps in the parameter space, the most efficient proposal distribution to have is a multi-variate Gaussian distribution.  The multi-variate jumps use a product of normal distributions in each eigendirection of the FIM, $\Gamma_{ij}$.  Here the Latin indices denote that this is the FIM on the 5-$D$ subspace of the intrinsic parameters.  The standard deviation in each eigendirection is given by $\sigma_{i} = 1/\sqrt{DE_{i}}$.  Here $D$ is the dimensionality of the search space.  As we are using the F-Statistic in our search phase,  $D=5$ and $E_{i}$ is the corresponding eigenvalue of $\Gamma_{ij}$.  The factor of $1/\sqrt{D}$ ensures an average jump of $\sim 1 \sigma$.  

The medium and large jumps were made from a variation of the same move.  In order to make a medium jump, we made a uniform draw in the five intrinsic parameters of $\pm 10\sigma$.  The large jumps came from a full range uniform draw on the intrinsic parameters, where the parameter range is defined by the priors for the input parameters.

\begin{figure}[t]
  \vspace{5pt}

  \centerline{\hbox{ \hspace{0.0in} 
    \epsfxsize=2.6in
   \epsffile{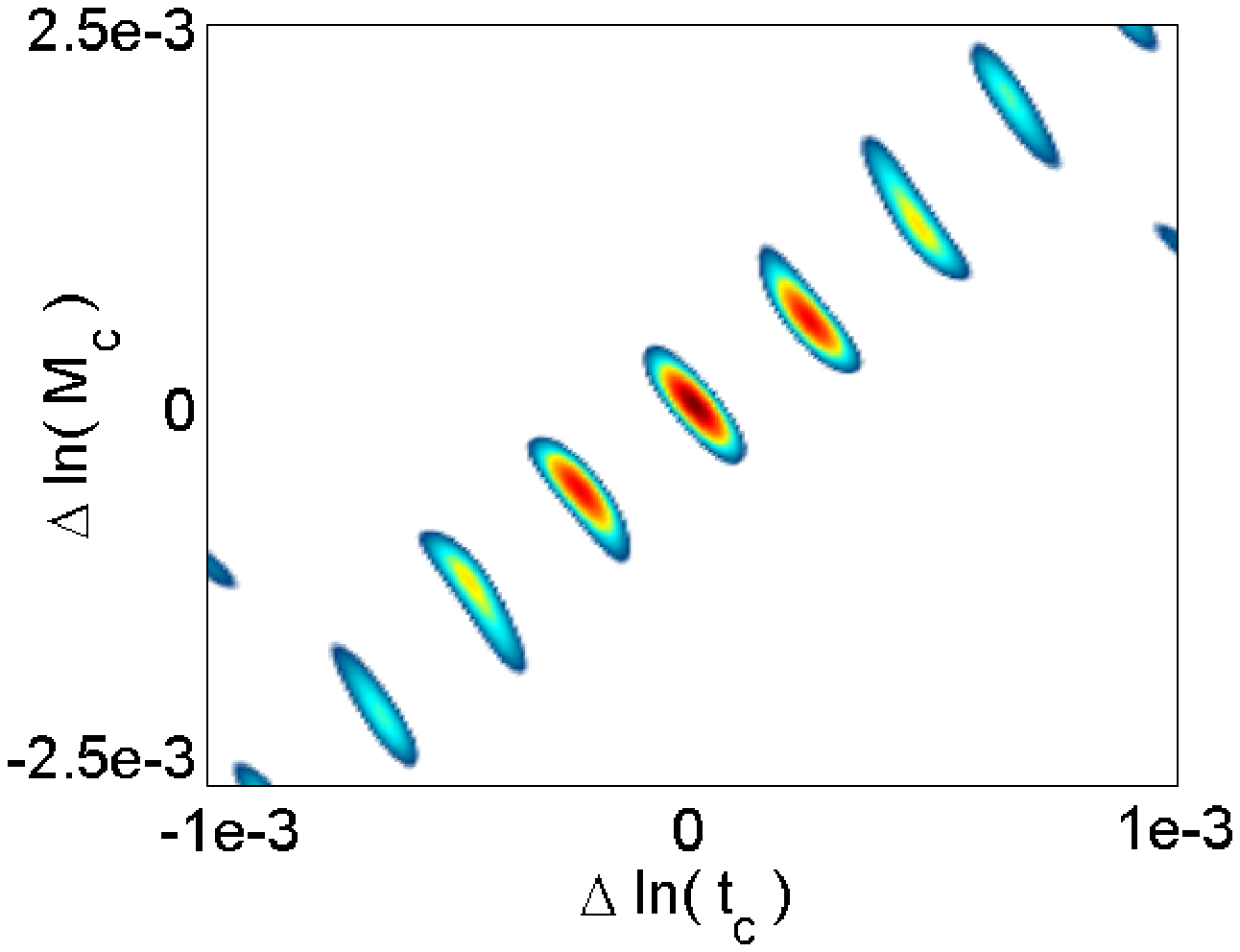}
    \hspace{0.05cm}
    \epsfxsize=2.7in
    \epsffile{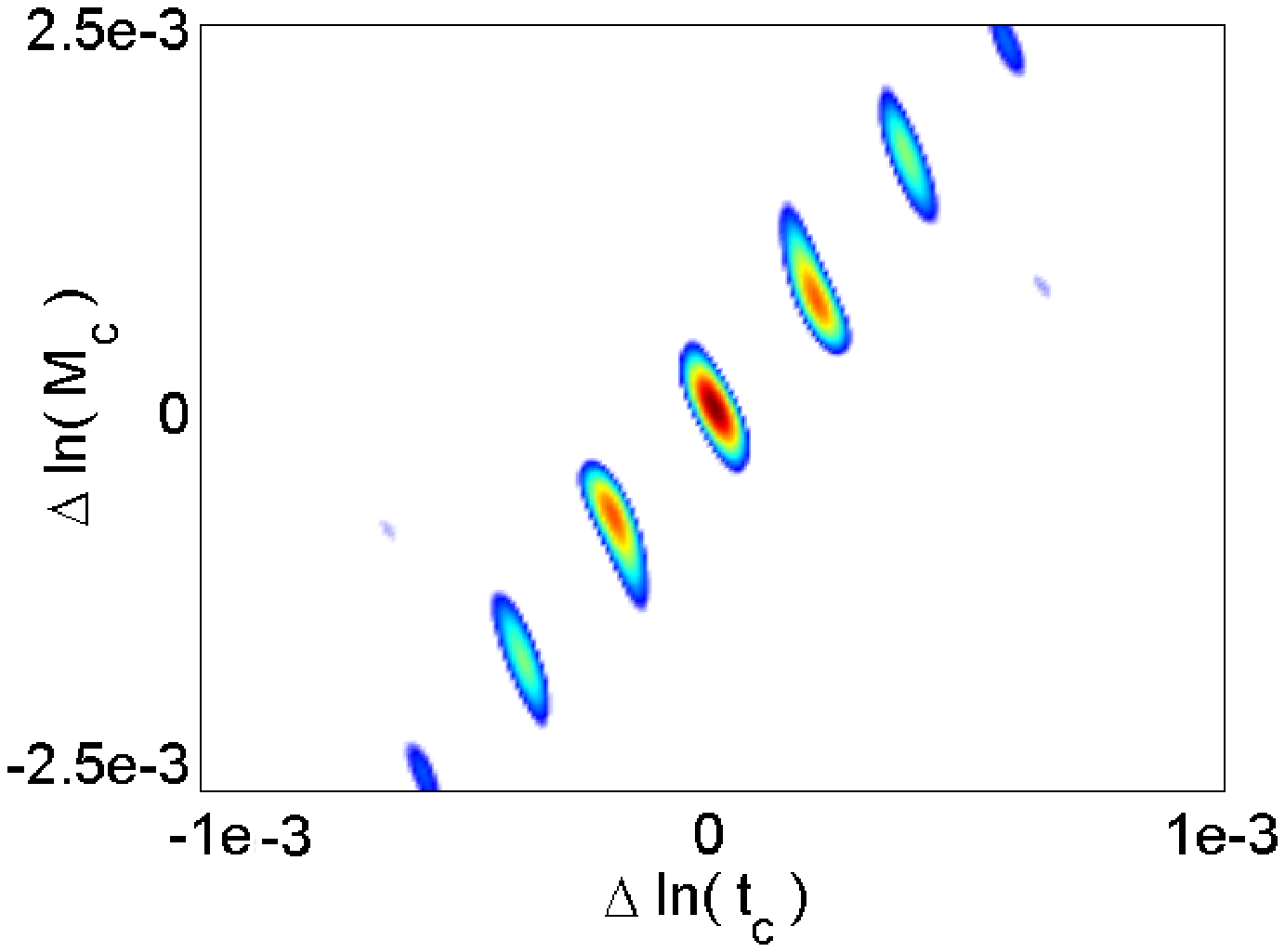}
    }
  }

  \vspace{5pt}
  \hbox{\hspace{1.in} (1 month) \hspace{1.9in} (2 weeks)}   \vspace{7pt}

  \centerline{\hbox{ \hspace{0.0in}
    \epsfxsize=2.7in
    \epsffile{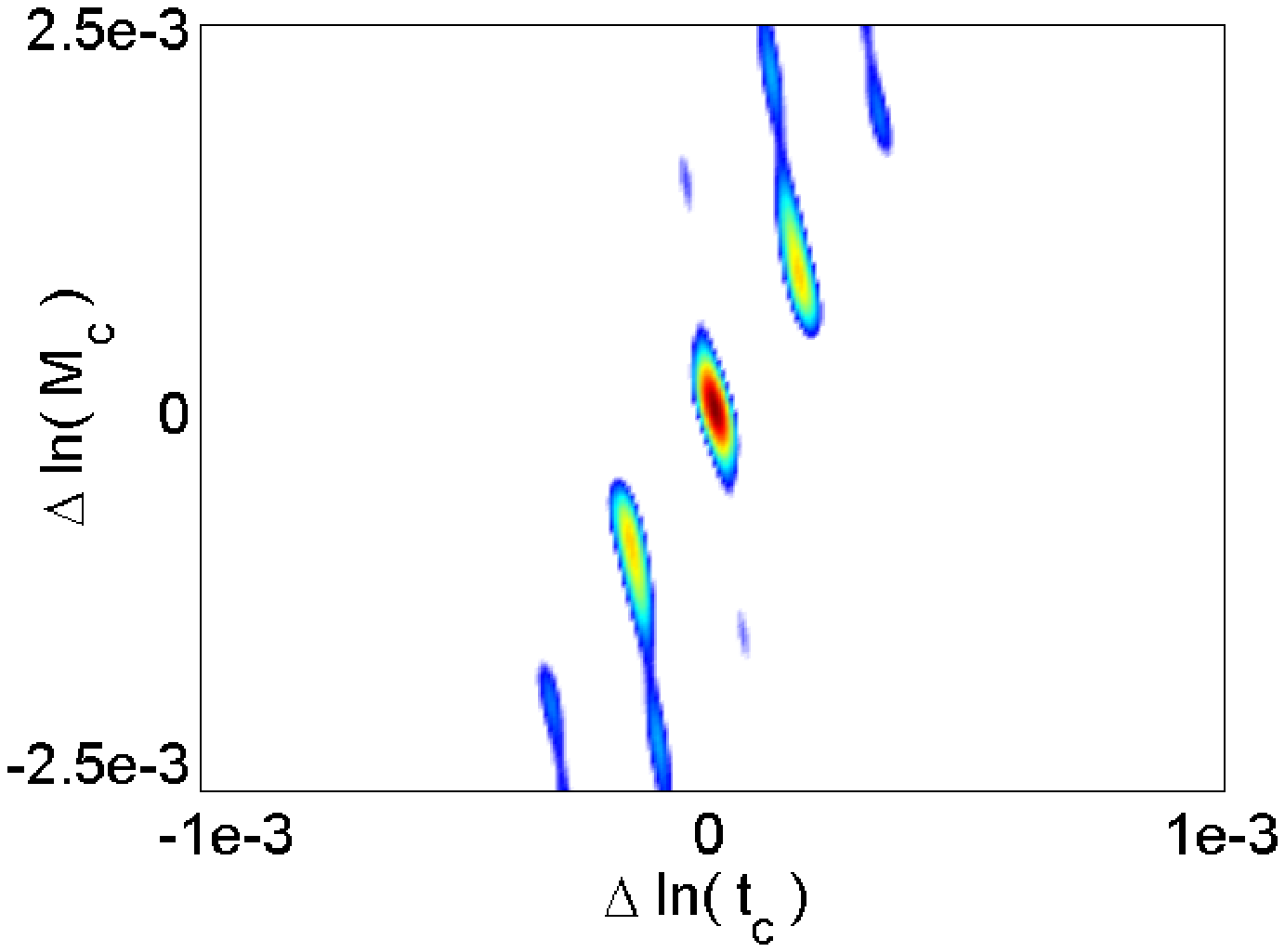}
    \hspace{0.05cm}
    \epsfxsize=2.7in
    \epsffile{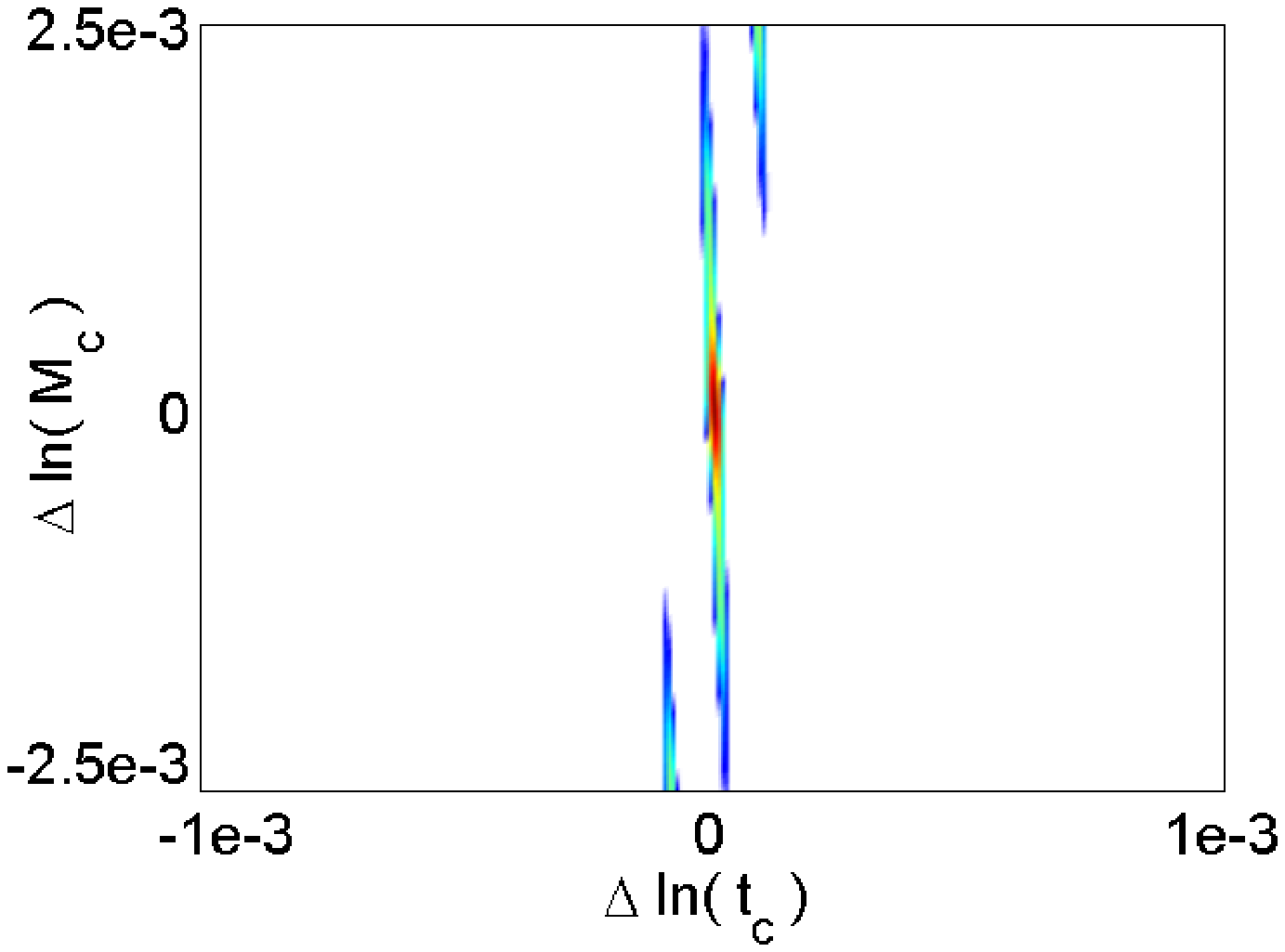}
    }
  }

  \vspace{5pt}
  \hbox{\hspace{1.in} (1 week) \hspace{2.in} (1 day)} 
  \vspace{5pt}

  \caption{ This Figure displays the evolution for the island chain of maxima on a $\ln\left(M_{c}\right)-\ln\left(t_{c}\right)$ slice through the Likelihood surface for coalescence times that exceed the observation time.}
  \label{fig:islandchains}

\end{figure}

Due to the quadrupole nature of the beam pattern functions of the detector, we also found it necessary to include a move that changed sky location by
\begin{equation}
\theta\rightarrow\pi-\theta\,\,\,\,\,\, , \,\,\,\,\,\,\,
\phi\rightarrow\phi\pm\pi.\nonumber
\end{equation}

By far the most important proposal distribution was a jump in the parameters $\{\ln M_{c}, \ln\mu, \ln t_{c}\}$.  While investigating the likelihood surface for MBHBs, we found that there were "island chains" of maxima running across the surface.  In Figure~\ref{fig:islandchains} we plot an example of these chains in the $\{\ln M_{c}, \ln t_{c}\}$ slice through the Likelihood surface.  Each cell of the plot represents a certain time to coalescence after the observation period has finished (we should mention here that for clarity, we set all points with a log-Likelihood less than zero equal to zero.  In reality, alongside each maximum, there is a minimum of equal or greater depth).  We can see that if the system coalesces a month after we finish observing there is a distinct chain of maxima almost bisecting the surface at an angle of $\pi/4$.  We should point out that while the islands are symmetric about the major axis of the central ellipsoid, the ellipsoids on one side of the center are flipped and rotated with respect to the ellipsoids on the other side of the image.  As we move through the cells from left to right, top to bottom, we notice that as the time of observation approaches the time of coalescence, two things happen.  Firstly, as expected, the central peak becomes more dominant on the Likelihood surface and the number of secondary peaks gets smaller.  Secondly, the entire chain begins to rotate as we acquire more information, breaking down the high correlation between various parameters.  As well as moving around the rest of the likelihood surface, we found that if we convinced the code to jump from island to island, we could greatly speed up the rate of convergence of the algorithm.  Therefore, our most important proposal distribution was to allow jumps in the eigendirections of these three intrinsic parameters according to 
\begin{equation}
x^{i}\rightarrow x^{i}+A\sqrt{\Delta x^{i}}
\end{equation}
Here, $A$ is a scaling constant and the small displacement in the eigendirection of each parameter, $\Delta x^{i}$, is given by
\begin{equation}
\Delta x^{i} = \sum_{j=1}^{5}\frac{V_{ij}^{2}}{E_{j}},
\end{equation}
where once again $E_{i}$ is the corresponding eigenvalue of the parameter, and $V_{ij}$ are the eigenvectors.

\subsection{Simulated annealing and $t_{c}$ maximization.}
As already stated, in a blind search, as we do not know where the actual signal lies, we could spend an eternity searching the likelihood surface.  To ensure a more acceptable movement in the chain, we use simulated annealing to heat the likelihood surface.  This smooths any bumps on the likelihood surface and spreads the width of the signal peak over a wider range.  This gives us a greater chance of moving uphill towards the summit of the peak.  In the definition of the likelihood, Equation~(\ref{eqn:likelihood}) we can see that there is a factor of 1/2 in the exponent.  We can replace this value with the parameter $\beta$, such that $\beta$ is given by
\begin{equation}
\beta = \left\{ \begin{array}{ll} \frac{1}{2}10^{-\xi\left(1-\frac{i}{T_{c}}\right)} & 0\leq i\leq T_{c} \\ \\ \frac{1}{2} & i > T_{c}  \end{array}\right.,
\end{equation}
where $\xi$ is the heat-index defining the initial heat, $i$ is the number of steps in the chain and $T_{c}$ is the cooling schedule.  

We should point out that one of the downsides of simulated annealing is that the initial heat factor and cooling schedule are dependent on the type of source we are looking for.  While the initial heat should be chosen high enough that the chain makes full range jumps in the parameters, it is harder to tell when we are overheating the surface.  We have found, by running many different chains, that the safest and easiest way to run the search is to chose a constant high value of $\beta$ for all searches.  However, this is wasteful, as low SNR sources do not require the same amount of heat as high SNR sources.  Generally we found that a cooling schedule of $10^{4}$ steps was more than sufficient if the initial heat had been chosen correctly.  It is important that the cooling schedule is not too short.  Even starting with a high heat, we found that if we cooled the chain too quickly it was likely to get stuck at a local minimum for a long time and not properly explore the parameter space.

The second measure we use to accelerate the convergence of the algorithm is to maximize over the time-of-coalescence, $t_{c}$, during the cooling phase of the chain.  We have found that this parameter is the most important one to find.  Once we have $t_{c}$, due to the fact that the parameters are highly correlated, the mass parameters are usually found very soon after.  Strictly, this is not a correct step to use as it assumes that LISA is stationary.  However, during the annealing phase we treat $t_{c}$ as being a quasi-extrinsic parameter and search over it separately.  Once the cooling phase has finished we stop this maximization.   The advantage of including this step in the search is that it manages to tie $t_{c}$ down to a restricted search range very quickly.  Our $t_{c}$ maximization is carried out using a modified F-Statistic search.  Using the usual Fourier domain $t_{c}$ maximization, the equations for the F-Statistic take on a slightly different form. 
This time instead of defining the four constants $N_{i}$, we define the matrix
\begin{equation}
N^{ij}=4\sum_{i=1}^{n}\sum_{j =1}^{4} \left(N^{ij}_{I}+N_{II}^{ij}\right),
\end{equation}
where $n$ is the number of elements in the waveform array.  The above equation describes the four constants at different time lags.  We can now solve for the time independent amplitudes
\begin{equation}
a_{ij} = \sum_{i=1}^{n}\sum_{j =1}^{4}\sum_{k =1}^{4}M_{jk}N^{ik}
\end{equation}
where the M-Matrix is the same as the one defined previously.  Inverting the M-Matrix, the F-Statistic is now a vector over different time lags.
\begin{equation}
{\mathcal F}_{i} = \sum_{i=1}^{n}\sum_{j =1}^{4}\sum_{k =1}^{4}M_{jk}N^{ij}N^{ik}
\end{equation}
We then search through this vector array to find the value of $t_{c}$ that maximizes the F-Statistic.  While it is possible to carry out a search without this maximization, it takes much longer to converge.  In fact, with $t_{c}$ maximization the algorithm has found the time to coalescence of the MBHB in as little as 10 steps of the chain.

\subsection{Frequency annealing.}
Due to limited computing power and the fact that we need to run the searches more than once to be sure we have found the correct source, we felt it necessary to improve the speed of the search algorithm.  For very high mass systems our simulated annealing code runs very quickly due to short waveforms.  However, as we start to look to lower mass systems, the computation time increases.  Therefore, instead of dealing with the entire search bandwidth, we choose to start off with a small subset of the data and increase the frequency range of the subset as the search progresses.  The main advantage of this method is that the initial part of the search is extremely fast as the array sizes are very small.  We have found in our runs that we usually have found the source by time we get to the costly part of the run.  This has the added advantage that the computationally expensive part of the algorithm is now more of a parameter refinement algorithm than a search algorithm.

The algorithm works as follows: we define a maximum search bandwidth frequency, $f_{max}$, which is a function of the lowest total mass in the priors.  We also define a bandwidth cut-off frequency, $f_{cut}$.  The initial upper cut-off frequency is chosen to be a multiple (at least 2, but in most cases 4) of the lower frequency cutoff of LISA.  Defining a growth parameter
\begin{equation}
B = \log\left(\frac{f_{max}}{a\,f_{cut}}\right)\,\,\,\,\,\,\,a\geq2, 
\end{equation}
we evolve the upper cut-off frequency according to 
\begin{equation}
f_{cut}= \left\{ \begin{array}{ll} 10^{-B\left(1-\frac{i}{N_{f}}\right)}f_{max} & f< f_{max} \\ \\ f_{max} & f \geq f_{max}  \end{array}\right.,
\end{equation}
where $i$ is the number of steps in the chain and $N_{f}$ is the total number of steps in the frequency annealing chain.  One of the main advantages of frequency annealing is that it acts as a kind of simulated annealing.  While the initial search is moderately accelerated by including simulated annealing, on the searches we have run, there is no difference in the rate of convergence by having the extra annealing included.  We therefore rid ourselves of having to make the best educated guess at what the initial heat factor should be.

The final refinement we use is to introduce a thermostated heat factor.  As the frequency annealing is a progressive algorithm (i.e. it does not see the full signal until very late in the run) we do not want it to get stuck on a secondary maximum.  To get around this we use a thermostated heat of 
\begin{equation}
\delta = \left\{ \begin{array}{ll} 1.0 & 0\leq SNR\leq \chi \\ \\ \left(\frac{SNR}{\chi}\right)^{2} & SNR > \chi  \end{array}\right. ,
\end{equation}
which ensures that once we attain a SNR of greater than $\chi$, the effective SNR never exceeds this value.  This thermostated heat means that the chain explores the parameter space more aggressively and as we get towards using the full template, there is enough heat in the system to prevent the chain from getting stuck.  We found in this analysis that it is sufficient to have $\chi=20$.  In this work, we ran the frequency annealing for 20,000 steps of the chain.  At the end of this search phase we take the current value of the thermostated head $\delta$ and cool to a heat of $\delta = 0.01$ over another 10,000 steps to find the Maximum Likelihood Estimates for the parameters.

\section{The Search for MBHBs.}\label{sec:results1}
We can divide the types of sources we are looking for into two distinct groups :  Those where we see coalescence and those where we do not.  Each case presents its own complications and must be treated slightly differently, especially when we come to mapping out the posterior distributions of the parameters and estimating the errors involved.  Due to this division, we treat each source type in turn.  

We should mention that Case A (Table~\ref{tab:paramsandpriors}) belongs to a class of astrophysically interesting sources.  There is a lot of evidence to suggest that modern galaxies formed from the coalescence of ``seed" galaxies in the distant universe.  In order to investigate whether or not LISA will be able to observe such sources, we have chosen a seed galaxy containing a MBHB with expected low masses at a high redshift of $z = 10$.   

\begin{figure}[t]
\begin{center}
\epsfig{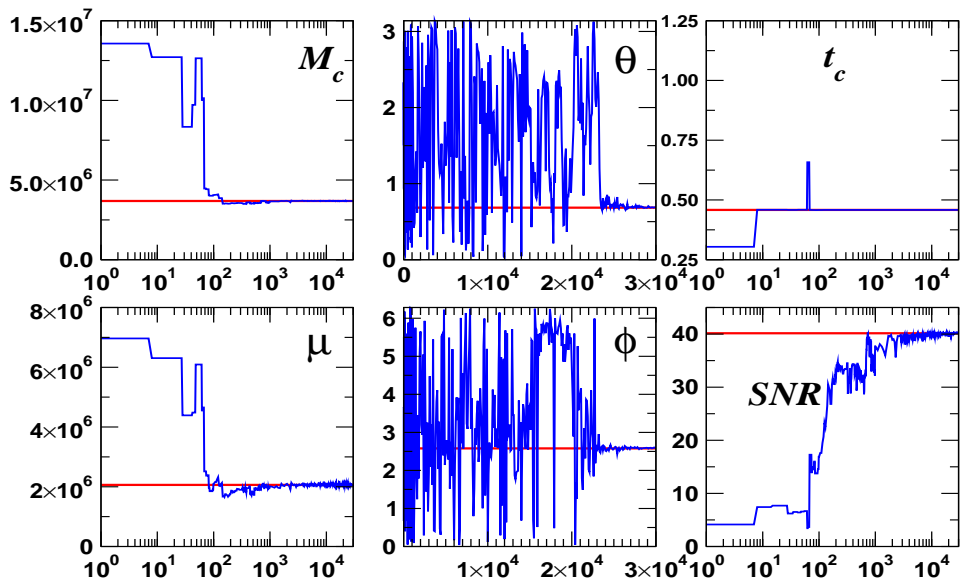}
\end{center}
\caption{A plot of the search chains for the case where we see the coalescence of the wave using simulated annealing.  The solid horizontal line in each cell denotes the true value of the parameters.}
\label{fig:search40sa}
\end{figure}

\begin{figure}[t]
\begin{center}
\epsfig{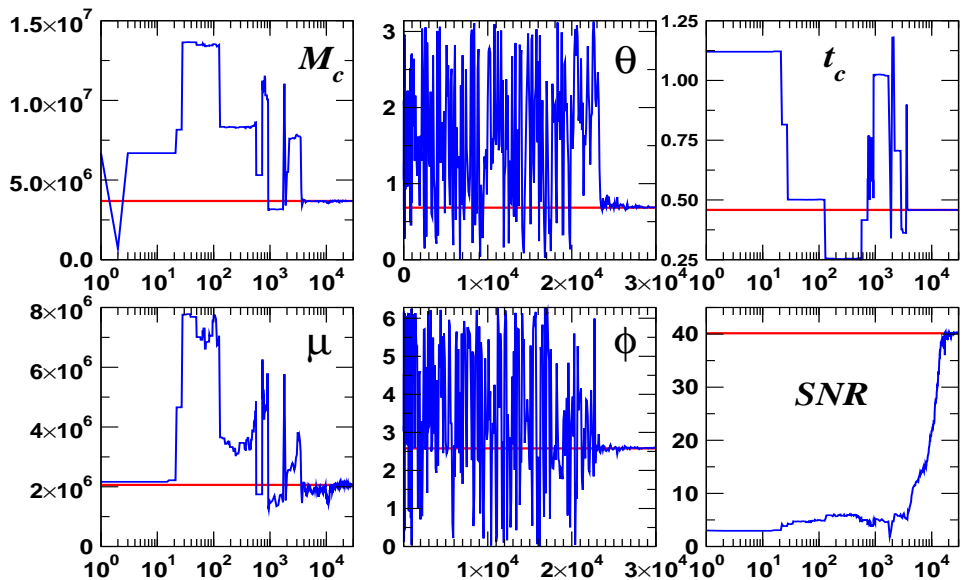}
\end{center}
\caption{A plot of the search chains for the case where we see the coalescence of the wave using a frequency annealing search.  The solid horizontal line in each cell denotes the true value of the parameters.}
\label{fig:search40fa}
\end{figure}


\subsection{Looking for signals where coalescence is observed.}
We produced simulated LISA data streams using the parameter values for Case A from Table~\ref{tab:paramsandpriors} and ran a search over $N=2\times10^{4}$ steps using both simulated and frequency annealing based searches.  In the case of the simulated annealing we chose a heat-index of $\xi = 2$ giving an initial heat of 100, while the  frequency annealing started with $f_{snip} = 4\times 10^{-5}$ Hz.  In both cases we chose a maximum bandwidth of $f_{max} = 2 f_{max}^{sig}(7M)$, where $f_{max}^{sig}(7M)$ is the maximum frequency of the injected signal at $7M$.  We then sampled with a cadence of $r_{samp}=2f_{max}$ for the simulated annealing. For the frequency annealing, while the input signals are generated at the above cadence, the templates are generated at a cadence of $2f_{snip}$ while $f_{snip} < f_{max}$.  When $f_{snip} = f_{max}$, we once again sample at $r_{samp}=2f_{max}$.  At the end of the search phase, we cool the chains to a heat of 0.01 over another $10^{4}$ steps.  This gives a reliable value of the Maximum Likelihood Estimate (MLE). 
  
In Figures~\ref{fig:search40sa} and \ref{fig:search40fa} we have plotted the search chains for the frequency and simulated annealing methods.  We should point out that because of the nature of the source, the algorithm does not find the sky positions until very late in the chain.  It is because of this that we have plotted the chains for $\theta$ and $\phi$ on a linear-linear scale, while all others are plotted on a log-linear scale.  In each cell, the solid horizontal line represents the true values of the parameters.

While both methods found the injected source, we can see from the figures that there are differences in the behaviour of each algorithm.  For example, we see a typical evolution in terms of the parameter convergence for the time to coalescence and both mass parameters.  With simulated annealing, the correct value of $t_{c}$ has been found in about 100 steps in the chain.  The two mass parameters then lock in in under a thousand steps.  We see that although the sky locations do not lock in until late in the search, we have recovered most of the SNR after 2000 steps in the chain.  On the other hand, it takes the frequency annealing almost 4000 steps before it closes in on $t_{c}$, $M_{c}$ and $\mu$.  This is due to the controlled growth in the available SNR that occurs when using frequency annealing. 

In Table~\ref{tab:snr40mle} we quote the MLEs at the end of the freezing phase for the frequency annealing chain (the values from the simulated annealing chain are almost identical), the standard deviation as predicted by the FIM at the injected parameter values and how close, in terms of multiples of the FIM standard deviations, the recovered values are to the injected values.  The retrieved values give a normalized overlap of ${\mathcal O}=0.9955$.  We should comment here on how well we retrieved the parameter values of the injected signal.  We mentioned earlier that we would expect that the parameter errors will have a Gaussian distribution around the correct value as given by the FIM.  If this was true, we would expect only $67\%$ of our answers to lie within $1\sigma$.  We attribute the better than expected performance to the anisotropic nature of the galactic foreground. The Fisher matrix analysis uses a sky averaged estimate for the galactic confusion noise, while the simulated galactic foreground is highly localized. MBHBs that are away from the galactic plane are less affected by the galactic foreground, and it is possible to beat the all-sky Fisher matrix error estimates.

It is also interesting to point out that there is a degeneracy in the solutions for the sky location.  In the LFA, the antennae response for LISA is almost perfectly quadrupole.  In the next section, where we look for a source where we do not see coalescence, we will include a chain that ended up on this degenerate solution.  We have found that when this happens, the difference in SNR between the two solutions is minimal.  In fact, for this particular case, one of the chains that found the alternate sky solution had a normalized overlap value of ${\mathcal O}=0.9954$, which is almost identical to the above solution.  In practice, this means we cannot call any sky solution that is out by $\pi - \theta$ in $\theta$, or $\pm\pi$ in $\phi$  a wrong answer.  In fact these solutions are just as valid as the ``real" values.  When we have run the chains multiple times with different starting seeds we observe that the chains quite happily interchange between the antipodal sky solutions.

\begin{table}[t]
\begin{center}
\begin{tabular}{|c|c|c|c|c|}\hline\hline
&  & & &   \\ 
& $\lambda^{X}$ & $\lambda^{MLE}$ & $\sigma^{FIM}_{\lambda^{X}}$ & $n\times\sigma^{FIM}_{\lambda^{X}}$\\ 
&  & & &   \\ \hline
&  & & &   \\ 
$M_{c}/M_{\odot}$ & $\,\,\,\,\,\,\,\,3.684932\times10^{6}\,\,\,\,\,\,\,\,$ & $\,\,\,\,\,\,\,\,3.686128\times10^{6}\,\,\,\,\,\,\,\,$ & $2.2255\times10^{3}$& -0.53\\ 
$\mu/M_{\odot}$ & $2.0625\times10^{6}$ & $2.06342\times10^{6}$ & $1.59\times10^{4}$ & -0.058\\
$\theta / rad$ & 0.6842 & 0.688477& $4.2397\times10^{-2}$& -0.1\\
$\phi / rad$  & 2.5791 & 2.59448& $6.0353\times10^{-2}$& -0.26\\
$t_{c}/yr$ & 0.458333 & 0.4583297& $7.429\times10^{-6}$& 0.49\\
&  & & &   \\ 
\hline\hline
\end{tabular}
\end{center}
\caption{This table compares the injected parameter values, the MLE for each parameter, the $1\sigma$ error predicted by the FIM at the injected values and the closeness of the MLEs to the injected values in multiples of the FIM $1\sigma$ error estimate for the source at $z=10$.}
\label{tab:snr40mle}
\end{table}

\begin{figure}[t]
\begin{center}
\epsfig{file=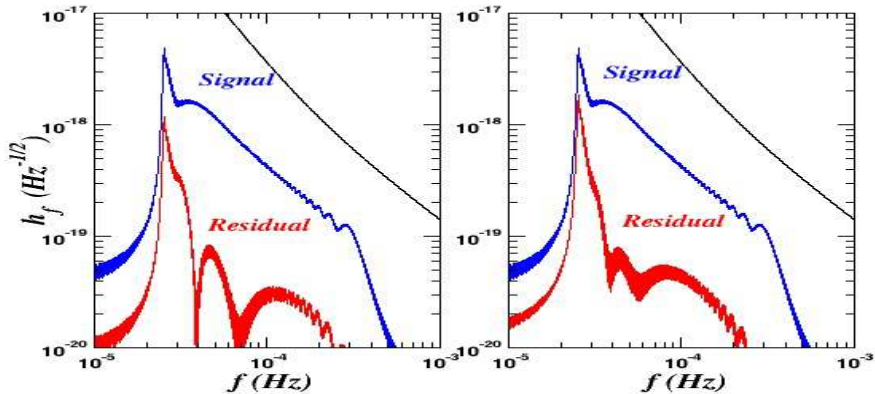, width=5in, height=2.5in}
\end{center}
\caption{A plot of the residuals for the source at $z = 1.5$.  The cell on the left shows the residual when we find the correct position on the sky.  The cell on the right shows the residual when we find the secondary solution on the opposite side of the sky.  As a point of reference, we have also included the rms noise level for the combined instrument and galactic foreground noise.}
\label{fig:search40res}
\end{figure}

Using the recovered parameters we plot the residuals of the signal, i.e. $|h_{inj}-h_{rec}|$, in Figure~\ref{fig:search40res}.  We also plot the residuals from a run where we found the degenerate sky solution.  We can see from the plot that besides a slight excess of power at high frequencies, there is not much difference in the two solutions.  

The main advantage of using the frequency annealing algorithm, is how quickly it locates the source as compared to using simulated annealing with the full signal.  In general the frequency annealing algorithm found the source in the space of minutes, while the simulated annealing algorithm took hours.  To compare the full run times (initial detection and parameter refinement), the codes were run on an AMD Athlon 2700+ with 1GB of memory.  For the case in question, the computing time was $\sim7.6$ hours to run the simulated annealing code and $\sim2.7$ hours to run the frequency annealing code.

\subsection{Looking for signals where coalescence is not observed.}
It turns out that the most interesting case is when we do not see the coalescence, which corresponds to LISA providing an early warning to the rest of the astronomical community.  The main problem for searching for signals with $T_{obs} < t_{c}$ is that the phase at coalescence is poorly determined.  The search chains for these type of systems reveal that $\varphi_{c}$ conducted virtually full range jumps over the length of the chain.  The indeterminacy of $\varphi_c$ has little effect on the search phase, but it does make it difficult to conduct a full 9-D exploration of the parameter posterior distribution function.  Using the values for Case B in Table~\ref{tab:paramsandpriors} we will treat both the search and posterior exploration phases in turn.

\begin{figure}[t]
\begin{center}
\epsfig{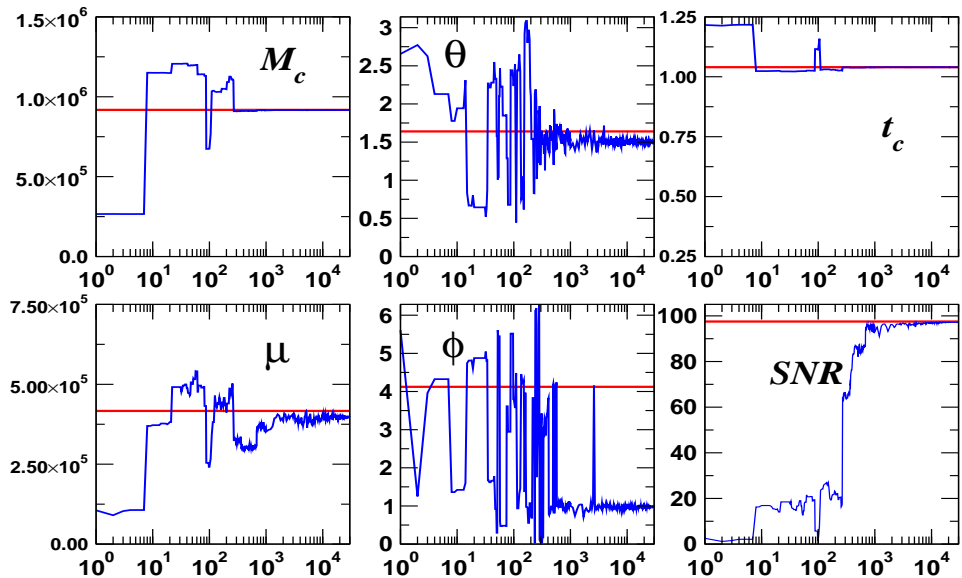}
\end{center}
\caption{A plot of the search chains for the case where we do not see the coalescence of the wave using simulated annealing.  The solid horizontal line in each cell denotes the true value of the parameters.  Note that in this case the chains find the alternative sky solutions.}
\label{fig:search97sa}
\end{figure}

\begin{figure}[t]
\begin{center}
\epsfig{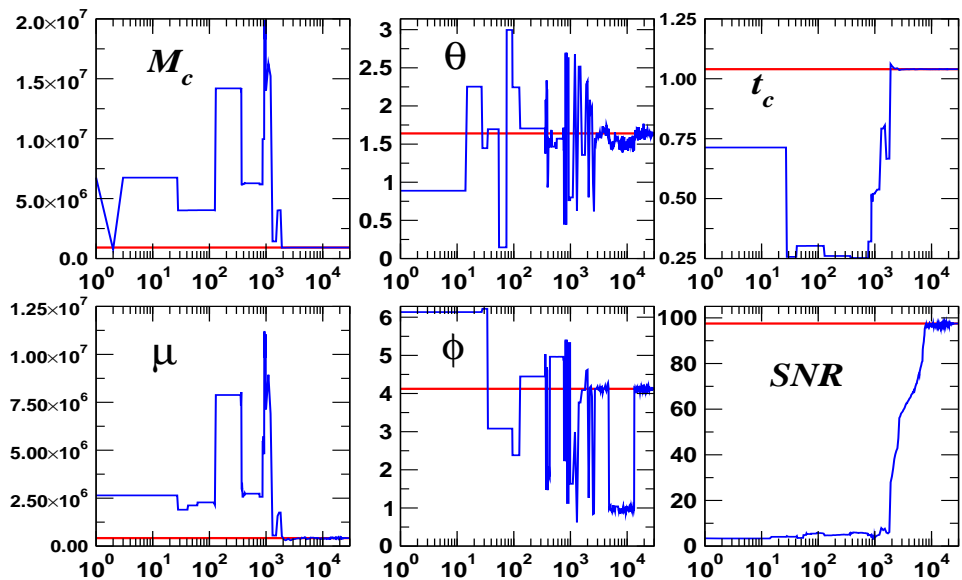}
\end{center}
\caption{A plot of the search chains for the case where we do not see the coalescence of the wave using a frequency annealing search.  The solid horizontal line in each cell denotes the true value of the parameters.}
\label{fig:search97fa}
\end{figure}

In Figures~\ref{fig:search97sa} and \ref{fig:search97fa}, we have again plotted the frequency and simulated annealing search chains.  We can see that both algorithms again find the injected source, however this time we have plotted a simulated annealing chain where we find the degenerate sky solution.  We can see from the bottom right cell in Figure~\ref{fig:search97sa} that there is no obvious reduction in SNR as compared to the frequency annealing search.  For the simulated annealing search, we started with a heat-index of $\xi = 1.7$ corresponding to an initial heat of almost 50.  The frequency annealing was once again started at $4\times 10^{-5}$ Hz.  We once again see that the important parameter to determine is $t_{c}$.  In the simulated annealing case, the chain has closed in on the correct vicinity of $t_{c}$  in about 10 steps of the chain and locks in properly after roughly 300 steps.  We also see that the chirp mass is also determined in about 300 steps.  It is interesting to see for the simulated annealing chain that as the chain locks in on the alternate sky position, it prevents the reduced mass from being more refined.  We can also see that due to the sky flips the chain does actually visit the correct sky location, but spends more of its time on the secondary solution.

In contrast, the frequency annealing chain takes longer to lock in on the true values ($\sim$2000 steps of the chain).  This is not surprising in this case.  As we previously said, we chose the initial frequency cut-off to be $4\times10^{-5}$ Hz.  For this signal, its initial frequency when it enters the LISA bandwidth is $4.33\times 10^{-5}$ Hz.  Therefore it takes a few hundred steps before the source even enters into the search, and then another couple of thousand steps before it accumulates enough enough SNR to resolve the key search parameters.  Note, that the first few thousand steps of the code where the source has been found takes only a few minutes to run.

In Table~\ref{tab:snr97mle} we quote the MLEs at the end of the freezing phase, the standard deviations as is predicted by the FIM and the closeness of the parameter fit in multiples of the FIM prediction.  These parameters, which are from the frequency annealing chain which found all of the parameters correctly, give a normalized overlap of ${\mathcal O}=0.99985$.  The MLEs from the simulated annealing chain that found the alternative sky solution gave an overlap of 0.9989, showing once again that it is virtually impossible to discern between the two solutions.

\begin{table}[t]
\begin{center}
\begin{tabular}{|c|c|c|c|c|}\hline\hline
&  & & &   \\ 
& $\lambda^{X}$ & $\lambda^{MLE}$ & $\sigma^{FIM}_{\lambda^{X}}$ & $n\times\sigma^{FIM}_{\lambda^{X}}$\\ 
&  & & &   \\ \hline
&  & & &   \\ 
$M_{c}/M_{\odot}$ & $\,\,\,\,\,\,\,\,9.17744\times10^{5}\,\,\,\,\,\,\,\,$ & $\,\,\,\,\,\,\,\,9.1774536\times10^{5}\,\,\,\,\,\,\,\,$ & $88.85983$& -0.011\\ 
$\mu/M_{\odot}$ & $4.16667\times10^{5}$ & $4.165817\times10^{6}$ & $2.3875\times10^{3}$ & 0.036\\
$\theta / rad$ & 1.6398 & 1.635726& $9.52755\times10^{-3}$& 0.43\\
$\phi / rad$  & 4.1226 & 4.118568& $9.46065\times10^{-4}$& 0.43\\
$t_{c}/yr$ & 1.04 & 1.0399988& $3.053062\times10^{-5}$& 0.037\\
&  & & &   \\ 
\hline\hline
\end{tabular}
\end{center}
\caption{This table compares the injected parameter values, the MLE for each parameter, the $1\sigma$ error predicted by the FIM at the injected values and the closeness of the MLEs to the injected values in multiples of the FIM $1\sigma$ error estimate for the source at $z=1.5$.}
\label{tab:snr97mle}
\end{table}

\begin{figure}[t]
\begin{center}
\epsfig{file=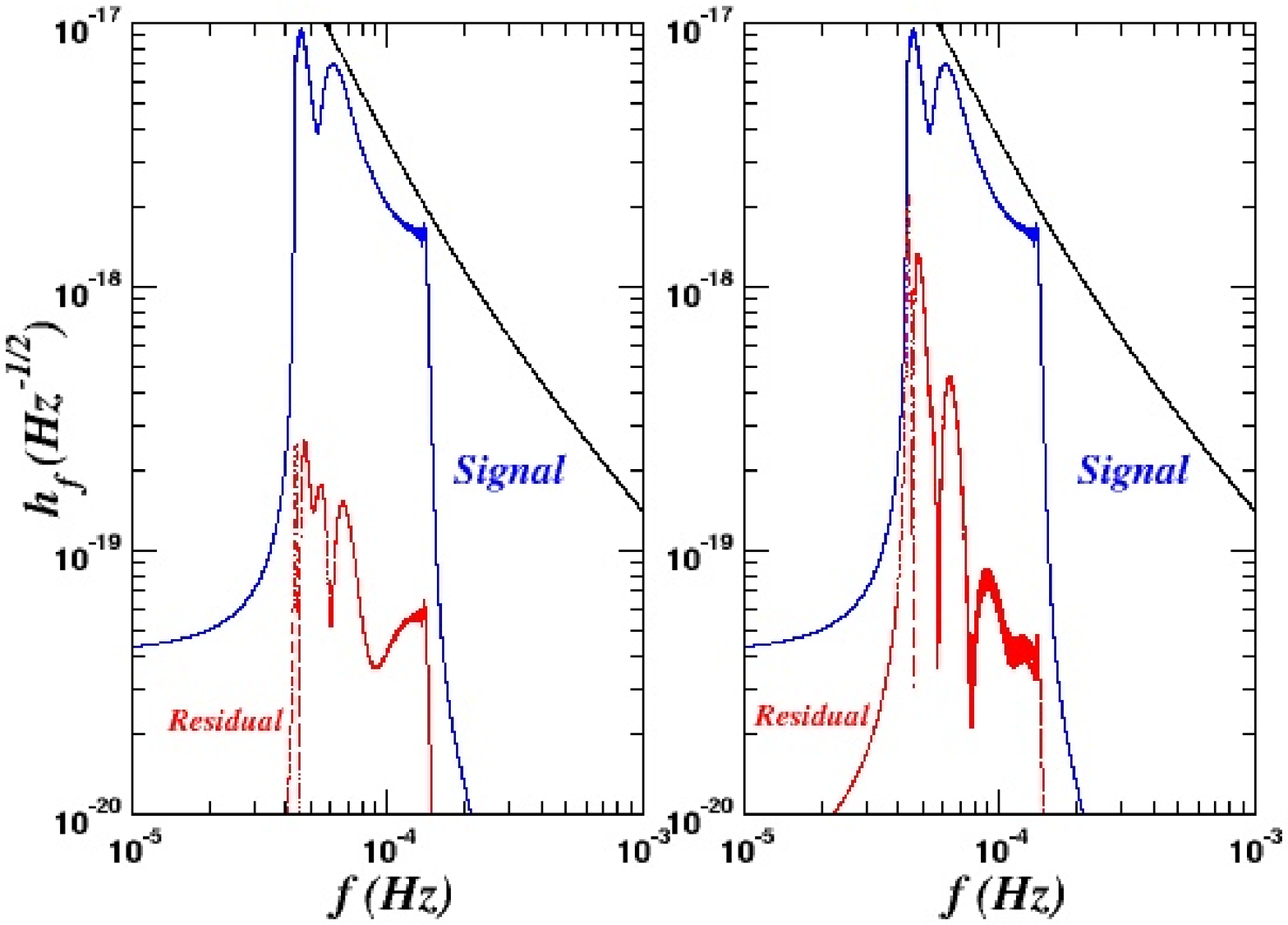, width=5in, height=2.5in}
\end{center}
\caption{A plot of the residuals for the source at $z = 10$.  The cell on the left shows the residual when we find the correct position on the sky.  The cell on the right shows the residual when we find the secondary solution on the opposite side of the sky.  As a point of reference, we have also included the rms noise level for the combined instrument and galactic foreground noise.}
\label{fig:snr40residuals}
\end{figure}

\section{Mapping out the posterior distribution functions using an adapted MCMC algorithm.}\label{sec:posteriors}

One of the problems for the MCMC method is that very strong correlations between parameters, can be detrimental to good mixing of the chain. In the case where we do see the coalescence, the correlations between the parameters is sufficiently mild that we can run a standard MCMC chain to explore the posterior distributions over the entire 9-$D$ parameter space~\cite{en2}.

\begin{figure}[t]
\begin{center}
\epsfig{file=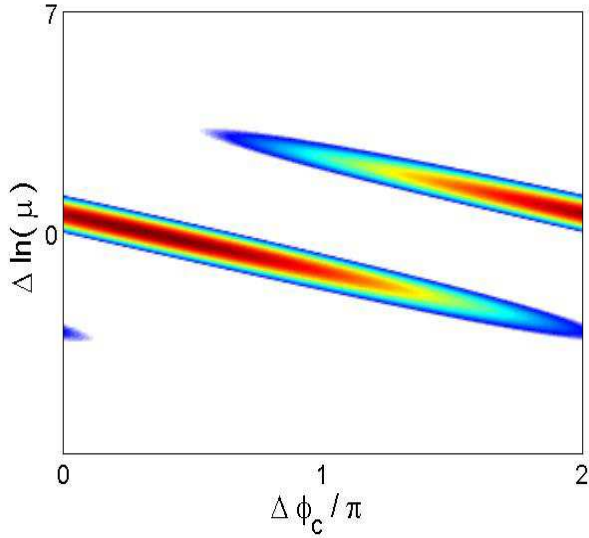, width=3.5in, height=3in}
\end{center}
\caption{A plot showing the elongated ellipsoid around the true values of $\ln(\mu)$ and $\varphi_{c}$ for the case where we do not see the coalescence of the binary.  If we think of the figure as representing an unfolded cylinder, we can see that the ellipsoid wraps itself around the cylinder multiple times.  The further we are out from coalescence, the greater the number of times the ellipsoid wraps itself around the cylinder.}
\label{fig:logLslice}
\end{figure}

However, the posterior exploration is more difficult prior to seeing the MBHB merger. In Figure~\ref{fig:logLslice} we have plotted a $\ln(\mu)-\varphi_{c}$ slice through the Likelihood surface for the case where do not see the coalescence of the MBHB.  We see that there is an extremely narrow extended ridge on the log-Likelihood surface.  To properly understand what is going on, we need to imagine the image as being an unrolled cylinder.  We can see that this ridge wraps itself around the cylinder multiple times.  In order to explore the posterior distributions properly, we need to travel up and down this ridge many times.  In fact, the further we are out from coalescence, the greater the number of times the ellipsoid wraps itself around the cylinder.  We can also see from the image the danger of either neglecting or fixing a constant value for $\varphi_{c}$.  Doing so would give incorrect errors for the other parameters as the chain will be confined to a small portion of the Likelihood surface.  

\begin{figure}[t]
\begin{center}
\epsfig{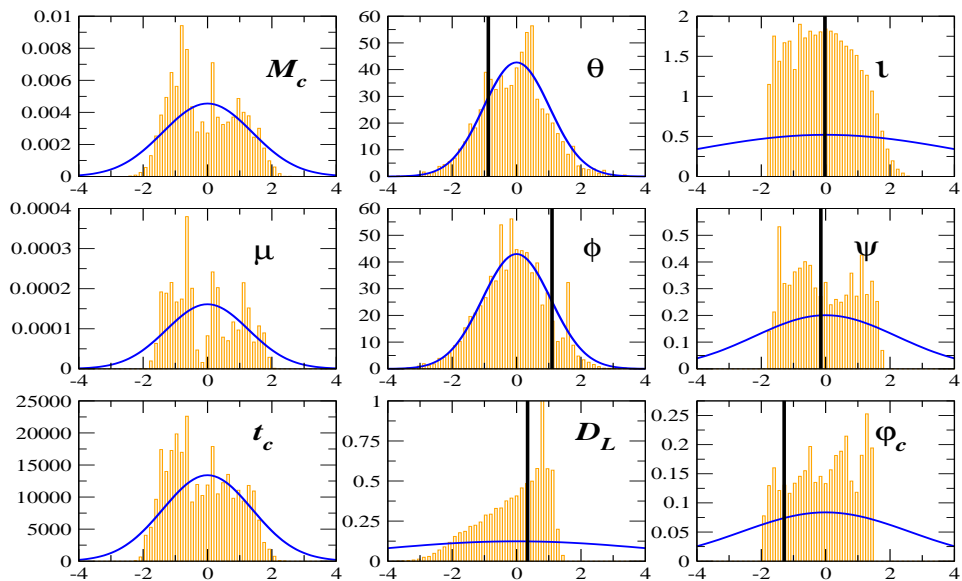}
\end{center}
\caption{A plot of the marginalized PDFs for a straightforward 9-D MCMC exploration of the waveform parameters.  Note that the means of the chain have been subtracted and the values scaled by the square root of the variances of the chains.  The solid line in each cell denotes the correct value.  In the cases where there is no solid line, this indicates that the chain has begun a random walk which has resulted in the mean of the chain being at a distance of $\geq 4\sigma$ from the correct answer.}
\label{fig:9dpost}
\end{figure}

To explore this further we first ran a full 9-D search using a normal proposal distribution to try an map out the posterior distribution functions (PDFs).  The search was run for $0.5\times10^{6}$ steps with an initial period of $10^{4}$ steps where the chain was run with a gentle simulated annealing with an initial heat of about 20.  This initial period is to allow the parameters to modify themselves slightly after the transition from a 5-D search to a 9-D exploration.  We should note that if the starting parameter values in the MCMC are slightly away from the correct values (i.e 10's of sigmas), they {\em will} migrate to better values, but without the simulated annealing it just takes longer for this to happen.  In Figure~\ref{fig:9dpost} we have plotted the marginalized PDFs against the prediction from the FIM at the mean of the chain.  The solid line in each cell denotes the correct value.  In the cases where there is no solid line, this indicates that the chain has begun a random walk which has resulted in the mean of the chain being at a distance of $\geq 4\sigma$ from the correct answer.  Note that histograms have had the mean values subtracted and the values are then scaled by the square root of the variances from the chains.  The Markov chains used to produce these histograms had very high auto and cross correlations, and we estimated the chains would need to be hundreds of times longer to ensure convergence to the correct posterior distribution. This may explain the mismatch between the PDFs and the Fisher matrix predictions, though it is also possible that the Fisher matrix predictions are unreliable due to the low SNR and the highly correlated source parameters.

There are a number of methods that could be used to improve the mixing of the chains. One approach would be to use adaptive MCMC techniques such as delayed rejection, or pilot chains that produce rough maps of the PDF, which can then be used to fashion more effective proposal distributions. Yet another approach is to observe that the search chains were little affected by the indeterminacy of $\varphi_c$ since the F-statistic eliminates this parameter from the search. On the other hand, the F-statistic also eliminates three other parameters that might like to include in our PDFs. One way around this is to use a mini F-Statistic that only eliminates two parameters, and thus allows us to implement a 7-D MCMC.  The mini F-Statistic can be derived by writing Equation~\ref{eqn:strainFS} in the form
\begin{equation}
h(t) = \frac{A_{+}}{D_{L}} \cos(\Phi) F^{+} + \frac{A_{\times}}{D_{L}} \sin(\Phi) F^{\times},
\end{equation}
where
\begin{eqnarray}
A_{+} &=& \frac{2Gm\eta}{c^{2}}x(t) \left(1+\cos^{2}\iota\right),\nonumber \\ A_{\times}&=&-\frac{4Gm\eta}{c^{2}}x(t) \cos\iota.
\end{eqnarray}
Expanding out the detector response in terms of $\varphi$ and $\varphi_{c}$ we get
\begin{eqnarray}
h(t) &=& \frac{\cos(\varphi_{c})}{D_{L}}\left[A_{+}F^{+}\cos(\varphi)-A_{\times}F^{\times}\sin(\varphi)\right]\nonumber\\
&+&\frac{\sin(\varphi_{c})}{D_{L}}\left[A_{+}F^{+}\sin(\varphi)+A_{\times}F^{\times}\cos(\varphi)\right].
\end{eqnarray}
On closer inspection we can see that the square bracket terms in the above expression correspond to the responses $h(t;\varphi_{c}=0)$ and $h(t;\varphi_{c}=\pi/2)$ respectively, with the luminosity distance set to unity.  We can then write the mini F-Statistic in the form
\begin{eqnarray}
h(t) & =&  \frac{\cos(\varphi_{c})}{D_{L}}\,h(t; \varphi_{c}=0, D_{L}=1)
 + \frac{\sin(\varphi_{c})}{D_{L}}\,h(t;\varphi_{c}=\pi/2, D_{L}=1)\nonumber\\
     & = & \sum_{k=1}^{2}\,a_{k}A^{k},
\end{eqnarray}
where
\begin{equation}
a_{1} = \frac{\cos(\varphi_{c})}{D_{L}}\,\,\,\,\,\, , \,\,\,\,\,\,\,\,\,
a_{2} = \frac{\sin(\varphi_{c})}{D_{L}},
\end{equation}
and
\begin{equation}
A^{1} = h(t; \varphi_{c}=0, D_{L}=1)\,\,\,\,\,\, , \,\,\,\,\,\,\,\,\, A^{2} = h(t;\varphi_{c}=\pi/2, D_{L}=1).
\end{equation}
We then repeat the steps from the generalized F-Statistic until we have the numerical values for the quantities $a_{k}$. We then find the maximized values of $\varphi_{c}$ and $D_{L}$ using
\begin{equation}
\varphi_{c} = \arctan\left(\frac{a_{2}}{a_{1}}\right),
\end{equation}
\begin{equation}
D_{L} = \left[a_{1}^{2} + a_{2}^{2}\right]^{-1/2}.
\end{equation}

\begin{figure}[t]
\begin{center}
\epsfig{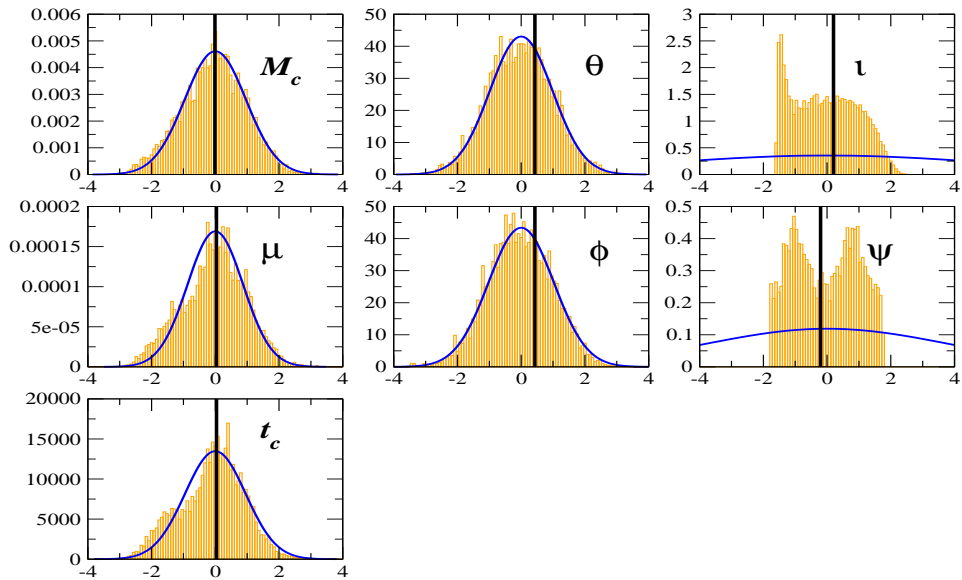}
\end{center}
\caption{A plot of the marginalized PDFs for a 7-D MCMC exploration of the waveform parameters where we have maximized over $\varphi_{c}$ and $D_{L}$.  Note that the means of the chain have been subtracted and the values scaled by the square root of the variances of the chains.  The solid line in each cell denotes the correct value.}
\label{fig:7dpost}
\end{figure}

In Figure~\ref{fig:7dpost} we plot the posterior distributions for the case where we apply the mini F-Statistic and search only over seven of the nine parameters.  The resulting Markov chains showed very little in the way of auto correlation, and the improved mixing yielded PDFs that are in good agreement with the Fisher matrix predictions for the five intrinsic parameters. The Fisher matrix predictions for the PDFs of the polarization and inclination angles are less accurate, but still reasonable.

\section{Searching for multiple Black Hole Binaries.}\label{sec:msmbh}
In this section we investigate the extraction of MBHBs when there is more than one system in the data stream.  Taking the two black hole binaries that we have already searched for individually, we injected both signals into the data stream.  While the optimal method would be to search for both MBHBs at the same time, in this instance we carry out the search iteratively.  Our plan is to subtract each source we find from the data set before commencing the search for the next source.  Once again we use a frequency annealing method to extract the systems.  While in practice one would use a global thermostated heat, our main objective here was to repeat the results of the individual searches when we had the two MBHBs present.  Therefore, we used the same thermostated heat factors for the individual sources in the two systems search as we did in the single MBHB search. 

\begin{figure}[t]
\begin{center}
\epsfig{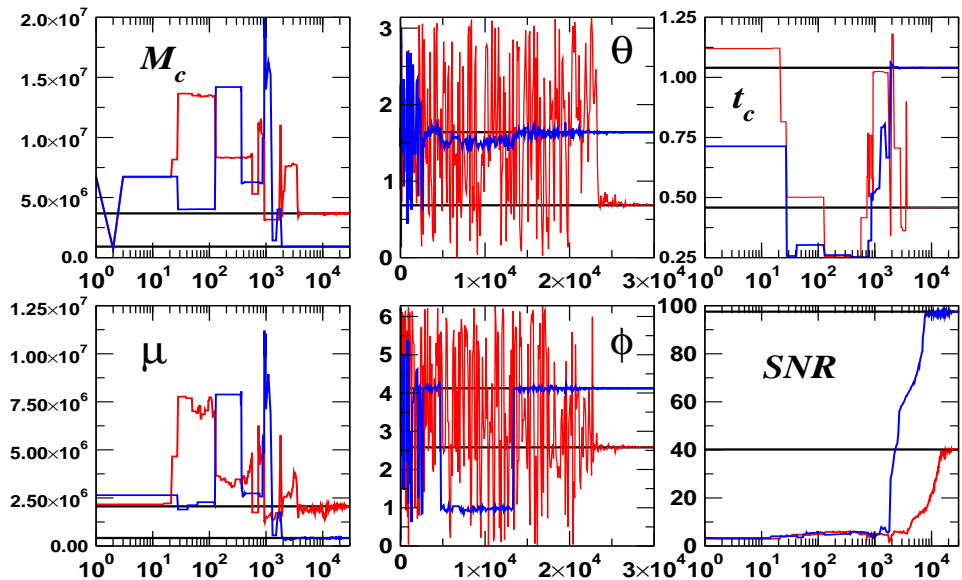}
\end{center}
\caption{A plot of the search chains for the case where we search for both MBHBs using a frequency annealing search.  The solid horizontal line in each cell denotes the true value of the parameters.}
\label{fig:2smbhsearch}
\end{figure}

The search for the two MBHBs was carried out multiple times with both frequency and simulated annealing methods, and in all cases the algorithm extracted the high SNR source first.  In none of our runs did the algorithm decide to go after the low SNR source first.  In Figure~\ref{fig:2smbhsearch}  we present the search chains for the two MBHB search.  Once again, the initial search phase was run for 20,000 steps, with an additional freezing period of 10,000 steps to extract the MLEs for the parameter values.  We can see from the search chains that there is almost no difference between the chains for the individual searches and the chains for the double MBHB search.  We should not be very surprised by this as the normalized overlap between the two sources is just -0.0203.

\begin{figure}[t]
\begin{center}
\epsfig{file=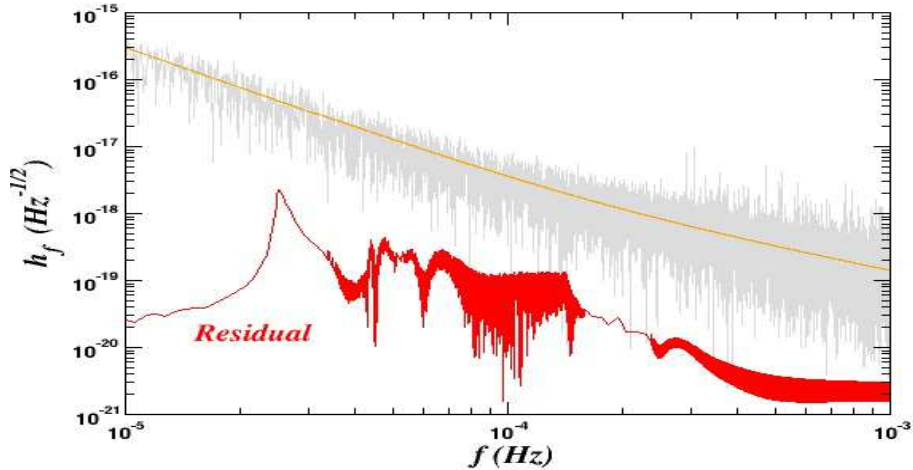, width=5in, height=3.in}
\end{center}
\caption{A plot of the noise residual after the subtraction of the two MBHBs.}
\label{fig:noiseres}
\end{figure}

The important question that has not been answered elsewhere is what effect will subtracting the MBHBs have on the galactic foreground.  In Figure~\ref{fig:noiseres} we plot the residuals of the noise, i.e. the difference between the original combination of instrumental and galactic foreground noise before the injection of the two MBHB sources, and the instrumental and galactic foreground noise after we have searched for and subtracted the two MBHBs.  We can see from the plot that there is minimal contamination of the galactic binaries and instrument noise.  This implies that we can safely extract the MBHB sources from the data without affecting the search for galactic binaries.  This supposition is currently being tested in an upcoming publication. 

\section{Conclusion.}
In this paper we have presented a new algorithm based on frequency annealing called Metropolis-Hastings Monte Carlo (MHMC).  We have introduced a galactic foreground of approximately 26 million individually modelled sources on top of instrument noise to give a realistic representation of the noise in the LISA output.  We chose our sources so that in one instance coalescence was seen and in another it was not, in one instance the source was quite close and bright, and in the other the source was quite far away and dim.  We demonstrated that both frequency and simulated annealing are successful in finding MBHBs in the LISA data stream even when no attempt is made to subtract the galactic foreground.  The main advantage that frequency annealing has over simulated annealing is that it is a more robust search algorithm.  In terms of real time analysis it finds the source faster than simulated annealing due to the fact that initially the sampling rates and array sizes are small.  Overall, the total runtime for the frequency annealing is also much shorter than the run time for simulated annealing allowing us to run many chains over again.

Once we have found the source, the next step is to map out the posterior distribution functions for the source parameters.  Due to the fact that the error ellipsoid in parameter space is so elongated and narrow, it is very difficult to get a straightforward MCMC chain to carry out an exploration of the parameter PDFs with reasonable acceptance rates and mixing properties.  This problem is amplified when we do not see the coalescence of the waveform.  We showed that good mixing of the Markov chains could be achieved by using a mini F-statistic that extremized over the phase at coalescence and the luminosity distance.

Finally, we investigated the case where there were two MBHBs in the data stream at the same time.  Using an iterative frequency annealing search we extracted both black hole binaries.  The extraction of each binary had no significant effect on the remaining binary nor on the noise in each detector channel. 

This work shows that LISA should be able to detect the inspiral of MBHBs out to the distant reaches of the universe.  Because of the nature of the source, we will also be able to detect multiple sources in the same data stream without effecting the galactic foreground.

\pagebreak
\appendix

\end{document}